\documentclass[aps,pra,twocolumn]{revtex4-1}
\usepackage[usenames]{color}
\newcommand{\comment}[1]{}
\usepackage{graphicx}
\usepackage{epsfig}
\usepackage{amsmath}
\usepackage{mathrsfs}
\newcommand{\be}{\begin{align}}
\newcommand{\ee}{\end{align}}
\newcommand{\bra}[1]{\langle {#1} |}
\newcommand{\ket}[1]{| {#1} \rangle}
\newcommand{\expect}[1]{\langle {#1} \rangle}
\newcommand{\ketn}[1]{ {#1} \rangle}
\newcommand{\bo}{{\bf \Omega}}
\usepackage{color}
 
\newcommand{\sA}{{\cal A}}
\newcommand{\bz}{{\bf z}}

\begin{document}
\title{Quantum rotor theory of spinor condensates in tight traps}
\date{\today}
\author{Ryan Barnett, Hoi-Yin Hui, Chien-Hung Lin,  Jay D. Sau, and S. Das Sarma}
\affiliation{Joint Quantum
Institute and Condensed Matter Theory Center, 
Department of Physics, University of Maryland, College
Park, Maryland 20742-4111, USA}

\begin{abstract}
In this work, we theoretically construct exact mappings of
many-particle bosonic systems onto quantum rotor models.  In
particular, we analyze the rotor representation of spinor
Bose-Einstein condensates.  In a previous work \cite{barnett10} it was
shown that there is an exact mapping of a spin-one condensate of fixed
particle number with quadratic Zeeman interaction onto a quantum rotor
model.  Since the rotor model has an unbounded spectrum from above, it
has many more eigenstates than the original bosonic model.  Here we
show that for each subset of states with fixed spin $F_z$, the
physical rotor eigenstates are always those with lowest energy.  We
classify three distinct physical limits of the rotor model: the Rabi,
Josephson, and Fock regimes.  The last regime corresponds to a
fragmented condensate and is thus not captured by the Bogoliubov
theory.  We next consider the semiclassical limit of the rotor problem
and make connections with the quantum wave functions through use of
the Husimi distribution function.  Finally, we describe how to extend
the analysis to higher-spin systems and derive a rotor model for the
spin-two condensate.  Theoretical details of the rotor mapping
are also provided here.
\end{abstract}
\maketitle

\section{Introduction}

The behavior of macroscopic systems of multicomponent bosons under
suitable constraints can often be greatly simplified through a quantum
rotor description. Within the context of condensed matter physics, the
most widely appreciated example is the celebrated Josephson model
\cite{josephson62,anderson67}.  This model provides an accurate
low-energy treatment of two superconductors linked by an insulating
barrier \cite{tinkham04}.  The treatment of the full many-particle
system reduces to a model with two canonically conjugate variables:
the relative particle number and phase between the two superconducting
regions.

Bose-Einstein Condensates composed of atoms with internal spin or
pseudospin degrees of freedom, the so-called spinor condensates, offer
another arena where such rotor mappings are highly useful.  Roughly
speaking, recent experimental work investigating the dynamics of
spinor condensates can be divided into two categories.  The first
category of experiments focuses on the complex interplay between
spatial and spin degrees of freedom resulting from spinor condensates
in larger traps \cite{sadler06, vengalatorre08,klempt09,
vengalattore10,klempt10, kronjager10}.  These experiments investigate
the dynamics after a quantum quench, which involves the proliferation
of topological defects.  The second catergory of experiments are
performed in tight traps where the spatial degrees of freedom are
unimportant
\cite{stenger98,chang04,schmaljohann04,chang05,widera05,mur-petit06,gerbier06,liu09a,liu09b}.
Such experiments have focused on the coherent spin dynamics after
preparing the system in a particular manner.  The rotor description is
useful  when the spatial degrees of the condensate can be
neglected, and thus is particularly relevant to the second class of experiments.

In an early theoretical work on spinor condensates it was shown that the
ground state of the antiferromagnetic condensate in tight traps
involves large spin correlations and can be considered to be a
condensate of singlet pairs of spin-one atoms \cite{law98}.  However,
such ``fragmented'' states \cite{ho00,mueller06} are known to be
extremely delicate and for most experimental situations are typically
better described by a broken symmetry state which is captured by the
classical Gross-Pitaevskii theory \cite{ho98,ohmi98,barnett09}.
Nevertheless the intriguing properties of the fragmented condensates
in the single-mode regime have motivated a considerable amount of
further theoretical work \cite{koashi00, ashhab02, diener06, chang07,
cui08, zhai09}.

In this paper we will revisit this problem by employing an exact rotor
mapping.  The mapping, which was carried out by some of us in a
previous work \cite{barnett10}, maps an antiferromagnetic spin-one
condensate in an external field onto a quantum rotor model of a
particle under an external field constrained to the unit sphere
\cite{barnett10}.  Since this mapping is exact, and not a low-energy
theory, it treats all possible phases of the spin one condensate 
on an equal footing.  Roughly speaking, states described by the
Gross-Pitaevskii Equation (GPE) correspond to rotor states that are
well localized in position.  On the other hand, states that are
delocalized over the sphere (e.g. the condensate of singlet pairs of
atoms) cannot be described by the GPE but are contained within this
rotor treatment.  We will provide in-depth analysis of the model, and
discuss its distinct physical regimes.  We will also describe its
semiclassical limit which has a clearer physical interpretation than the
GPE and elucidate the semiclassical behavior of the rotor
wave functions for appropriate parameter regimes.  We will also
describe how to extend the mapping to systems with larger spin.
In that sense, the current work is a generalization and extension of
Ref.~\cite{barnett10}.

The paper is organized as follows.  In Sec.~\ref{Sec:doublewell}, for completeness, we
consider the simplest nontrivial example of bosons in a double-well potential and
map this system onto a quantum rotor model.  We arrive at a result
first obtained in Ref.~\cite{anglin01} but we use a method that can be
generalized to systems with more components i.e. higher spins.  In
Sec.~\ref{Sec:spinone} we move on to the more complex case of a
spin-one condensate in the single mode regime and overview the rotor
mapping originally derived by some of us in Ref.~\cite{barnett10}.  In
Sec.~\ref{Sec:correspondence} we consider the correspondence between the
eigenvalues of the original bosonic problem (which has a finite
spectrum for fixed particle number) and the rotor model (which has an
unbounded spectrum from above).  In Sec.~\ref{Sec:spectra} we consider
in more detail the spectrum of the rotor model, and establish three
distinct physical limiting cases, namely the Rabi, Josephson, and Fock
regimes.  We provide analytic expressions for the low-lying
spectrum for these cases.  In Sec.~\ref{Sec:semiclassical} we
consider the semiclassical limit of the rotor model. Here
we discuss recent experiments on $^{23}$Na dynamics in terms
the semiclassical phase space.  We then connect 
the quantum mechanical wave functions in the Rabi and Josephson
regimes to the semiclassical phase space using  a generalization
of the Husimi distribution function \cite{husimi40,mahmud05}.
In Sec.~\ref{Sec:ext} we consider extending the
rotor mapping to larger component systems, focusing on the example
of the spin-two condensates.  Finally, in Sec.~\ref{Sec:conclusion} we
conclude with a summary.

\section{Bosons in a double-well potential}
\label{Sec:doublewell}

In this Section, we consider the simplest nontrivial
case of bosons in a double-well potential which is described by the
dimer Bose-Hubbard model.  This archetypal model has been studied
extensively \cite{milburn97,smerzi97,anglin01,leggett01,mueller06} and
has also been used to experimentally oberserve the so-called
self-trapping effect \cite{albiez05}.  In the interesting work of
Anglin \emph{et al.} \cite{anglin01} it was shown that the dimer
Bose-Hubbard model can be exactly mapped onto a two-dimensional
quantum rotor model.  Below we will derive their main result, using a
different formalism which allows a more direct generalization to higher
dimensional rotor systems which will be considered in following
sections.  We clarify our notations and lay out the main theoretical
framework in this section by considering the double-well case first.

Our starting point is the Bose-Hubbard dimer model which describes
bosons in a double-well potential with repulsive interactions
\begin{align}
\label{Eq:doublewell}
H= -J (a_1^\dagger a_2 + a_2^\dagger a_1) + \frac{1}{2} U n_1(n_1-1) + 
\frac{1}{2} U n_2(n_2-1).
\end{align}
Here $a_{1}^\dagger$ and $a_{2}^\dagger$ create bosons in the left and
right wells respectively, $n_{\alpha}= a_{\alpha}^\dagger a_\alpha$ is
the particle number operator, $J$ is the hopping, and $U$ is the
on-site repulsion.  It is often instructive to use the amplitude-phase
representation of the bosonic operators.  That is, we can write
$a=\sqrt{n_\alpha} e^{i\theta_\alpha}$ and impose the commutation
relation $[n_\alpha,\theta_\beta]=i \delta_{\alpha\beta}$.
Inserting these relations into Eq.~(\ref{Eq:doublewell}) and expanding
to leading order in the total particle number $N=n_1+n_2$ (which is
taken to be fixed) leads to the well-known Josephson model
\cite{josephson62, anderson67}
\begin{align}
\label{Eq:Jos}
H_{\rm Jos}= -JN \cos(\theta) +U n^2
\end{align}
where $\theta=\theta_1 - \theta_2$ and $n=( n_1-n_2)/2$ so that the
two operators in this equation are canonically conjugate: $[n,\theta]
= i$.  The spectrum of the Josephson model can be seen to agree with
the original double-well model Eq.~(\ref{Eq:doublewell}) in the
large-particle number limit.
 
In the work of Anglin \emph{et al.} \cite{anglin01}, it was shown that
such a mapping can be made exact, and thus will reproduce the spectrum
of Eq.~(\ref{Eq:doublewell}) for arbitrary particle number.  Their
derivation used a method akin to the Bargmann phase-space
representation of bosonic operators \cite{bargmann61}.  
Here we will derive their central result through a different method.
To start, we define the states 
 \begin{align}
 \ket{\bo_N}&= \frac{1}{\sqrt{2^N N!}} \left( a_1^\dagger e^{i\theta}
+ a_2^\dagger e^{-i\theta} \right)^N \ket{0}\\ &=\frac{1}{\sqrt{N!}}
\left( \bo \cdot {\bf b}^\dagger \right)^N \ket{0}
 \end{align}
 where $\bo=(\cos(\theta),\sin(\theta))$ is a real two-component
vector on the unit circle and the ``Cartesian'' bosonic operators
$b_{x}$ and $b_y$ are defined to be $b_x =
\frac{1}{\sqrt{2}}(a_1+a_2)$, $b_y = \frac{-i}{\sqrt{2}}(a_1-a_2)$.
These states can be shown to form an overcomplete basis.  For
instance, the fragmented state $(a_1^\dagger)^{N/2}
(a_2^\dagger)^{N/2}\ket{0}$ can be seen to be an equal weight
superposition of these states over the unit circle \cite{mueller06}.
Therefore, an arbitrary state $\ket{\Psi}$ in the bosonic Hilbert
space can be expressed in terms of a superposition over the states
$\ket{\bo_N}$ with weight factor $\psi(\bo)$
 \begin{align}
 \label{Eq:expand}
 \ket{\Psi} = \int d\Omega \ket{\Omega_N} \psi(\bo).
 \end{align}
Note that due to the overcompleteness, this relation does not uniquely
determine $\psi(\bo)$.  The approach of the mapping is to find a
Hamiltonian ${\cal H}$ acting in the ``rotor'' space such that
\begin{align}
\label{Eq:method}
\int d\Omega \left( H \ket{\Omega_N}  \right) \psi(\bo) = \int d\Omega \ket{\Omega_N}  \left( {\cal H} \psi(\bo) \right).
\end{align}
Then the rotor Schrodinger equation ${\cal H} \psi = i
 \partial_t \psi$ is a \emph{sufficient} condition for the
bosonic Schrodinger equation to be satisfied (we will work in units
where $\hbar=1$ unless otherwise stated).

In obtaining ${\cal H}$, we use the gradient operator ${\bf \nabla}$
on the unit circle, which  in terms of $\theta$
is $\nabla_x = -\sin(\theta)\partial_\theta$ and $\nabla_y = \cos(\theta) \partial_\theta$.  These derivatives satisfy
the geometrically intuitive relations
\begin{equation}
\nabla_\alpha \Omega_\beta = \delta_{\alpha \beta} - \Omega_\alpha \Omega_\beta
\end{equation}
(for a discussion see  Appendix~\ref{A1}).
With this, it can be seen that quadratic operators acting on $\ket{\bo_N}$ can be written as
\begin{equation}
\label{Eq:rule}
b_\alpha^\dagger b_\beta \ket{\bo_N} = \Omega_\beta (\nabla_\alpha + N \Omega_\alpha) \ket{\bo_N}.
\end{equation}

In terms of the Cartesian operators, the double-well 
Hamiltonian (up to a constant offset) is 
\begin{align}
H= -J (b_x^\dagger b_x - b_y^\dagger b_y) + \frac{U}{4} (i b_x^\dagger b_y - i b_y^\dagger b_x)^2.
\end{align}
We can now use the relation in Eq.~(\ref{Eq:rule}) to find
\begin{align}
(b_x^\dagger b_x - b_y^\dagger b_y) \ket{\bo_N} = \left( N\cos(2\theta) -\sin(2\theta) \partial_\theta \right) \ket{\bo_N}
\end{align}
and
\begin{align}
(i b_x^\dagger b_y - i b_y^\dagger b_x)^2\ket{\bo_N} = L_{xy}^2 \ket{\bo_N}
\end{align}
where $L_{xy} = -i \Omega_x \nabla_y + i \Omega_y
\nabla_x=-i \partial_\theta$.  
The effective rotor Hamiltonian can then
be obtained by
inserting these relations into Eq.~(\ref{Eq:method}),
and integrating by parts.  One finds
\begin{equation}
{\cal H} = \frac{1}{4} U n^2 - J (N+2)\cos(2\theta) - i J \sin(2\theta) n
\end{equation}
where $n=i\partial_\theta$.  While the operator ${\cal H}$ has a real
spectrum it is 
not Hermitian due to its last term.  
However, one can apply a similarity transform to render ${\cal H}$ Hermitian.  Specifically, defining  \cite{anglin01}
\begin{align}
{\mathscr H} = e^{\cos(2\theta) \frac{J}{U}} {\cal H} e^{-\cos(2\theta) \frac{J}{U}} 
\end{align}
and shifting $\theta \rightarrow \theta/2$ to compare with Eq.~ (\ref{Eq:Jos}) 
one finds
\begin{align}
\label{Eq:rotor2}
{\mathscr H} = U n^2 - J(N+1) \cos(\theta) + \frac{J^2}{U} \sin^2(\theta)
\end{align}
which is the main result.
Note that this reduces to Eq.~(\ref{Eq:Jos}) in the large-$N$ limit.
The additional terms in Eq.~(\ref{Eq:Jos}), however, serve to
make the spectrum of  the original double-well Hamiltonian
Eq.~(\ref{Eq:doublewell}) exactly match the eigenstates of this rotor model.

\section{Spin-one condensates in the single-mode regime}
\label{Sec:spinone}

We now move on to discuss the related, but more complex, problem of
the spinor condensate in the single mode regime under a magnetic
field.  Recently it was shown \cite{barnett10} that this system maps
onto a quantum rotor model under an external magnetic field.  Here we
will summarize this mapping.

Our starting point is a spin-one condensate in a trap that is
sufficiently tight such that it is a good approximation to take all of
the bosons to occupy the same spatial mode.  Under this approximation,
we can write the field operators for each spin state as 
\begin{equation}
\psi_\alpha({\bf r}) = \phi({\bf r}) a_\alpha 
\end{equation}
where $\alpha$ runs
from $-1$ to $1$.  The condensate profile satisfies
\begin{equation}
\int d^3 r | \phi({\bf r})|^2 = N
\end{equation}
where $N$ is the number of particles in the system.
This approximation,
commonly referred to as the single mode approximation, breaks down
when the condensate coherence length is larger than the condensate
size.   

The Hamiltonian for this system reads 
\begin{equation}
\label{Eq:H} 
H = \frac{g}{2N} F^2 - q a_0^\dagger a_0.
\end{equation} 
In this equation, ${\bf F}=a^\dagger_{\alpha} {\bf F}_{\alpha\beta}a_{\beta}$ is
the total spin operator where ${\bf F}_{\alpha \beta}$ are spin-one
matrices,  $g$
is the spin-dependent interaction, and $q$ is the quadratic Zeeman
shift due to an external magnetic field.  Taking a uniform
condensate density $\phi({\bf r}) = \sqrt{n_0}$, we can express 
$g$ in terms of microscopic parameters as
\begin{equation} 
g=\frac{4\pi
\hbar^2}{3m}(\bar{a}_2 - \bar{a}_0)n_0
\end{equation} 
where $m$ is the mass of the constituent atoms, and
 $\bar{a}_0$ and $\bar{a}_2$ are the scattering lengths.
We will focus on the case of antiferromagnetic interactions
for which $g>0$ as is the case for $^{23}$Na condensates.

As was done for the double-well problem in Sec.~\ref{Sec:doublewell},
it is useful to transform the bosonic operators to the Cartesian
basis, rewriting the operators as
$
b_x = -(a_1 - a_{-1})/\sqrt{2},
$
$
b_y  = (a_1 + a_{-1})/i\sqrt{2},
$
and
$
b_z = a_0.
$
Written in terms of these, the spin operator becomes 
${\bf F}=-i{\bf b}^\dagger \times {\bf b}$.  
We next define the overcomplete set of states as
\begin{align}
\ket{\bo_N}=\frac{1}{\sqrt{N!}} \left( \bo \cdot {\bf b}^\dagger \right)^N \ket{0}
\end{align} 
which are parametrized by a three-component vector on the unit sphere
$\bo$ (note that the analogous states in Sec.~\ref{Sec:doublewell}
were parameterized on the unit circle).  

The general mapping proceeds with the general method given above in
Sec.~\ref{Sec:doublewell}.  Namely, one writes a general bosonic wave
function as a superposition of states in the $\ket{\bo_N}$ basis, and
finds an operator $\cal{H}$ acting in the rotor Hilbert space which
satisfies Eq.~(\ref{Eq:method}) (where the integration is generalized to
the unit sphere).  The full derivation is given in
Ref.~\cite{barnett10} and thus we will only give the results here.
The rotor Hamiltonian corresponding to Eq.~(\ref{Eq:H}) is
\begin{equation}
\label{Eq:Hrnh}
{\cal H}  = \frac{g}{2N} L^2 -  q(N+3)\Omega_z^2 + q \Omega_z
\nabla_z
\end{equation}
where $L_\alpha$ is the angular momentum operator and $\nabla_\alpha$
are the gradient operators on the unit sphere.  In the spherical
coordinate representation, $\nabla_z= -\sin(\theta) \partial_\theta$.
This can be brought to the more intuitive  Hermitian form by applying
a similarity transformation.  In particular, defining 
${\mathscr H} = e^{-S} {\cal H} e^{S}$ where
$S=\frac{qN}{4g}\cos(2\theta)$,
we find
\begin{equation}
\label{Eq:Hr} 
{\mathscr H}= \frac{1}{2I} L^2 + V(\theta)
\end{equation} 
where  $I=N/g$ is the moment of inertia, and 
\begin{equation}
\label{Eq:V}
V(\theta) = q\left(N+\frac{3}{2}\right)
\sin^2(\theta) + \frac{q^2N}{8 g}\sin^2(2\theta)
\end{equation}
is the external potential.  The spectrum of this Hamiltonian exactly
matches that of Eq.~(\ref{Eq:H}). As was described in \cite{barnett10}
one must retain only the eigenstates of this rotor model which
are symmetric under inversion: $\psi(\bo)=\psi(-\bo)$.  However,
since the operator which projects into this subspace of states commutes with
the rotor Hamiltonian, this imposes no additional conceptual or technical difficulty.

Since $\phi$ does not appear in the potential $V$ in
Eq.~(\ref{Eq:Hr}), one notes that this rotor model has azimuthal
symmetry.  This symmetry essentially reduces the model to a
one-dimensional system which considerably simplifies computations.
One should note, however, that we did not rely on this symmetry in the
derivation and it will not be present for more general
couplings.   In Appendix \ref{A2} we provide a rotor
mapping for a more general coupling.

\section{Correspondence of the Rotor and bosonic eigenvalues}
\label{Sec:correspondence}

There are subtleties that arise due to the fact that the rotor model
Eq.~(\ref{Eq:Hr}) has an unbounded spectrum from above, while the
spectrum of the original bosonic problem for fixed particle number $N$
if finite.  As is clear from the mapping, an eigenstate of the rotor
model $\psi$ is a sufficient condition for an eigenstate of the
bosonic Hamiltonian $\ket{\Psi}$.  That is, given $\psi$, one can
construct the bosonic eigenstate through
\begin{equation}
\label{Eq:transform}
\ket{\Psi} = \int d\Omega \ket{\Omega_N} \psi(\bo).
\end{equation}
Here for simplicity we are taking $\psi$ to be an eigenstates of the
non-Hermitian rotor model ${\cal H}$ so that we do not need to include
factors of $e^{S}$.  Because the spectrum of the
bosonic Hamiltonian $H$ is bounded, the only possibility is that many
of the rotor eigenstates get transformed to $\ket{\Psi}=0$ through
Eq.~(\ref{Eq:transform}), noting that this trivially satisfies the
bosonic Schrodinger equation.  Following Ref.~\cite{anglin01} we will
refer to the rotor eigenstates which transform to $\ket{\Psi} \ne 0$
as ``physical'' and those that transform to $\ket{\Psi} = 0$ as
``unphysical''.

We next ask if all of the eigenstates of the bosonic spectrum are
included in the rotor description.  For instance, the pathological
case of all the rotor eigenstates mapping to $\ket{\Psi}=0$ is not
\emph{a priori} ruled out.  Another question that arises regards the
ordering of the unphysical and physical eigenstates.  In particular,
is there an energy cutoff below which all eigenstates are physical and
above which eigenstates are unphysical?  We will show that there is an
affirmative answer to both of these questions.

\begin{figure}
\includegraphics[width=3.5in]{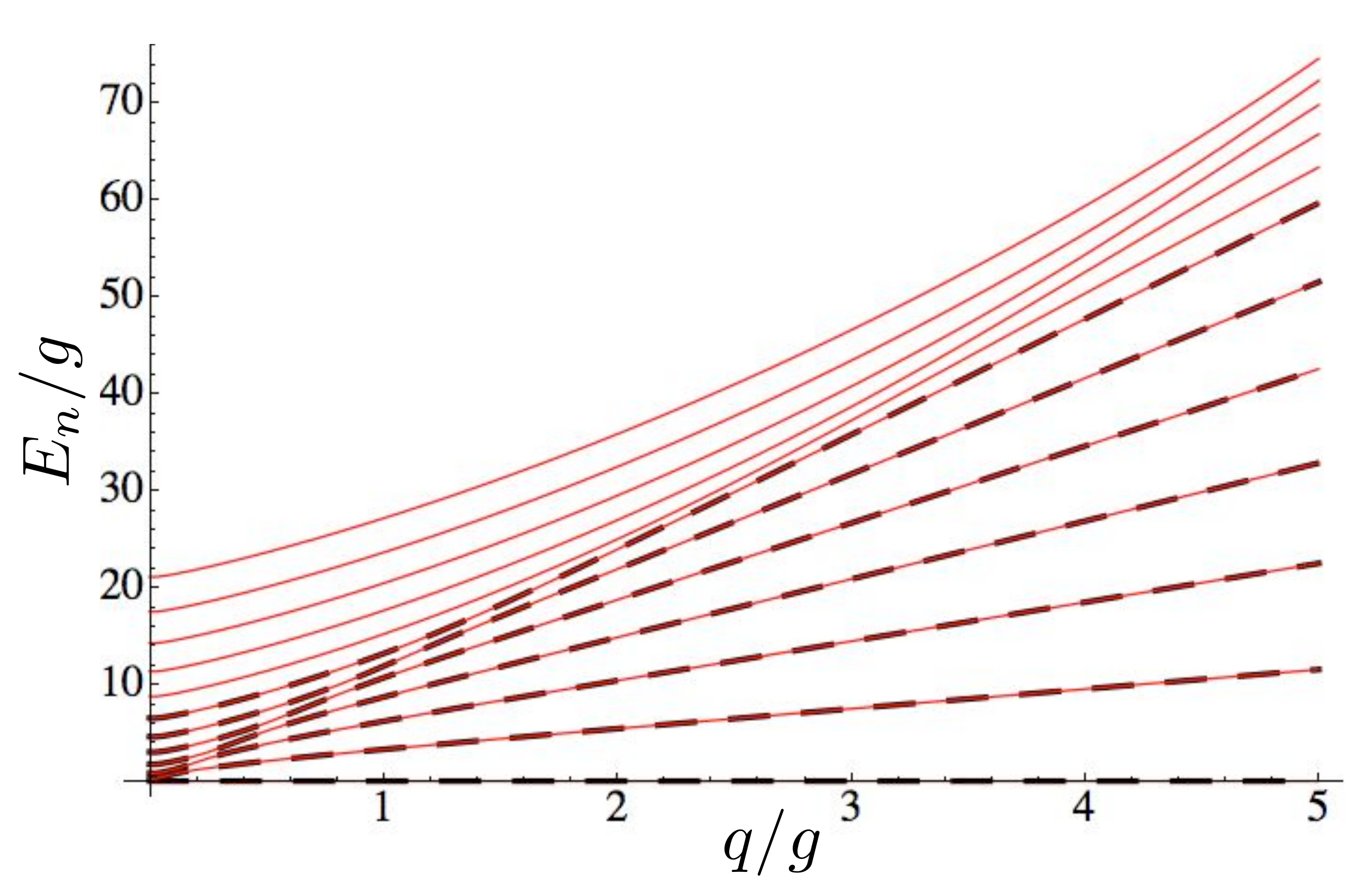}
\caption{Black dashed lines: the eigenvalues $E_n$ of the spinor
bosonic Hamiltonian Eq.~(\ref{Eq:H}) for $N=12$ particles for fixed
$F_z=0$ as a function of quadratic Zeeman field $q$. Red solid lines:
the lowest $12$ eigenvalues of the rotor Hamiltonian Eq.~(\ref{Eq:Hr})
for fixed $L_z=0$ after the antisymmetric states for which
$\psi(\bo)=-\psi(\bo)$ are projected out.  The lowest (physical)
eigenvalues of the rotor model exactly agree with those of the bosonic
Hamiltonian.  For all points in the plot, the spectrum is shifted so
that the lowest eigenvalue has zero energy.}
\label{Fig:espec}
\end{figure}

For sufficiently small particle number $N$, the eigenspectrum of the 
bosonic Hamiltonian Eq.~(\ref{Eq:H}) can be numerically computed and
compared to the eigenspectrum of the rotor system Eq.~(\ref{Eq:Hr}).
The rotor Hamiltonian has azimuthal symmetry since the potential
$V$ appearing in Eq.~(\ref{Eq:V}) does not depend on the angle $\phi$.
Therefore, ${\mathscr H}$ commutes with $L_z$ and the eigenspectrum
for fixed values $L_z = m$ can be considered separately.  Similarly,
$F_z$ commutes with the bosonic Hamiltonian $H$, and we can compare
to the rotor model by fixing $F_z=-m$.  In Fig.~\ref{Fig:espec} the
eigenspectrum of both the rotor model and the bosonic model
are shown as a function of the quadratic Zeeman field $q$.  We take
the case of relatively small particle number $N=20$, and take fixed
$F_z=L_z=0$.  As can be seen, all of the eigenenergies of the bosonic
Hamiltonian are accounted for by the rotor model.  Furthermore, the
physical states of the rotor model are always lower in energy than the
unphysical states.  Similar behavior was found for other values of
fixed $L_z=m\ne 0$ which are not shown.

This behavior can be understood as follows.  For unphysical states
$\psi(\bo)$, it can be seen from Eq.~(\ref{Eq:transform}) that
$\bra{Y_{\ell m}}\ketn{\psi} = 0$ for all $\ell \le N$.  It can be
verified that the (non-Hermitian) rotor Hamiltonian defined in
Eq.~(\ref{Eq:Hrnh}) has the property
\begin{equation}
\label{Eq:nohopping} \bra{Y_{\ell m}} {\cal H} \ket{Y_{\ell' m'}}=0
\end{equation} 
for $\ell \le N$ and $\ell' > N$.  Suppose that we have
a rotor eigenstate which is unphysical for parameters $(q,g)$.  Then
the eigenstate at $(q+\Delta q, g)$ can be determined by first order
perturbation theory.  By doing so, one sees from
Eq.~(\ref{Eq:nohopping}) that if a state is unphysical at $q$ then the
same state will also be unphysical at $q+\Delta q$.  In the limit of
$q=0$, the rotor model becomes trivial.  Here the eigenstates are
simply spherical harmonics with eigenenergies $E_{\ell} = \frac{g}{2N}
\ell(\ell+1)$.  Furthermore, the lowest eigenstates for $\ell \le N$
are all physical while the higher eigenstates for $\ell > N$ are
unphysical in this limit.  We note that for fixed $L_z=m$, the rotor
Hamiltonian becomes one dimensional.  Thus there will not be any band
crossings \footnote{This can be shown by assuming two solutions to
Eq.~(\ref{Eq:Hr}) with the same energy.  It can be shown that the resulting
Wronskian vanishes and thus the two solutions are equal to each other
up to a multiplicative constant}.
From the perturbative argument above, we therefore
conclude that the higher energy states will always remain unphysical
and not mix with the lower energy physical states.

 \section{Eigenspectra of the Rotor Model}
 \label{Sec:spectra}

In this Section we will concentrate on the eigenspectrum of the
spin-one rotor hamiltonian given in Eq.~(\ref{Eq:Hr}).  We will give the
spectrum in particular limiting cases, and compare the results with
those the Bose-Hubbard Dimer problem.  

We consider how the
spectrum evolves as a function of $q$.  For large $q$, the potential
energy $V(\theta)$ in Eq.~(\ref{Eq:V}) serves to localize the wave function on the unit
sphere.  To obtain the energy levels, the angular momentum $L^2$ can
be expanded about the north pole so that ${\mathscr H}$ becomes a
two-dimensional harmonic oscillator.  When $1 \ll q/g$, the second
term in the potential energy $V(\theta)$ dominates
so that the energy levels are given by
\begin{equation}
\label{Eq:RabiE}
E_{(n_x, n_y)} = q (n_x + n_y ) 
\end{equation}
 where $n_x$ and $n_y$ are integers corresponding to the oscillator
modes in the $x$ and $y$ directions. These eigenstates can in fact 
be directly obtained  from the original bosonic hamiltonian Eq.~(\ref{Eq:H})
in the large-$q$ limit.

Next we consider the case of smaller $q$ where 
$1/N^2 \ll q/g \ll 1$.  For this case, the first term in the
potential energy is the most significant.  Here we can also expand the kinetic
energy about the north pole to obtain a harmonic oscillator
hamiltonian.  For this the energy levels read
\begin{equation}
\label{Eq:JosephsonE}
E_{(n_x, n_y)} = \sqrt{2 g q} (n_x + n_y ) 
\end{equation}
where, as in Eq.~(\ref{Eq:RabiE}), $n_x$ and $n_y$ are integers.
As shown in Appendix \ref{A3} it can be seen that the Bogoliubov
spectrum of Eq.~(\ref{Eq:H}) agrees with 
Eqns.~(\ref{Eq:RabiE}, \ref{Eq:JosephsonE}).

Finally we consider the case of vanishingly small magnetic field such
that $q/g \ll \frac{1}{N^2}$.  For this case the eigenfunctions are not
localized about the north pole.  The kinetic energy $\frac{1}{2I} L^2$
dominates the rotor model and the eigenstates are given simply
by
\begin{equation}
\label{Eq:FockE}
E_{\ell} = \frac{g}{2N} \ell (\ell + 1).
\end{equation}
Each of these energy levels has multiplicity $2\ell + 1$.  So that
the wave function has inversion symmetry, only even values of $\ell$
should be kept.  The ground state in this regime  
where the rotor is completely delocalized about the
unit sphere, in terms of the bosonic model, is the fragmented
condensate composed of singlet pairs of bosons.  However, 
due to the condition $q/g \ll 1/N^2$, in the thermodynamic limit any small magnetic
field will drive the system to a symmetry broken state which is 
described well by mean field theory \cite{ho98,ohmi98,barnett09}.  
This is the central difficulty in
experimentally realizing the singlet condensate.  We will address this
problem in more detail in Appendix.~\ref{A4}.

It is instructive to compare the above results with the dimer
Bose-Hubbard model.  This model has been analyzed and found to have
three distinct limits, namely the ``Rabi'', ``Josephson'', and
``Fock'' regimes using the terminology of Leggett
\cite{leggett98,leggett01}.  Using a method very similar to that used
above, the expressions for the energy eigenstates can be obtained in
these regimes from the rotor Hamiltonian in Eq.~(\ref{Eq:rotor2}).
Namely, for the Rabi regime where $N \ll J/U$ the last term in the
potential energy dominates and the spectrum is approximated by a
harmonic oscillator, after expanding about
$\theta=0$.  The Josephson regime occurs when the first term in the
potential energy dominates $1/N \ll J/U \ll N$.  Here the states are
also localized about $\theta=0$.  Finally, for $J/U \ll 1/N$ the Fock
regime is obtained where the rotor is delocalized over the unit
circle.  The Josephson Hamiltonian Eq.~(\ref{Eq:Jos}) correctly
describes the Fock and Josephson regimes, but cannot describe
the Rabi regime since a large-$N$ expansion is used to derive it.
In summary, the three possible regimes for the dimer Bose-Hubbard
model are
\begin{align}
\label{Eq:condition1}
 N \ll J/U \;& : {\rm Rabi} \\
1/N \ll J/U \ll N \; &:  {\rm Josephson} \\
J/U \ll 1/N \; &: {\rm Fock}.
\end{align}

It is clear that there is a close parallel between the above
described regimes for the dimer Bose-Hubbard model and those of the
spin-one condensate problem.  For this reason we will adopt the
terminology introduced in \cite{leggett98,leggett01} for the spinor
problem.  Namely, we will label the three regimes as
\begin{align}
\label{Eq:Rabi}
1 \ll q/g \;& : {\rm Rabi} \\
\label{Eq:Josephson}
1/N^2 \ll q/g \ll 1 \; &:  {\rm Josephson} \\
\label{Eq:Fock}
q/g \ll 1/N^2 \; &: {\rm Fock}.
\end{align}
For typical experimental situations (for example those describe in
Refs.~\cite{liu09a,liu09b}) $q \sim g$ and $N\sim 10^3-10^5$
which places the system in either the Rabi or Josephson regimes.
For these cases, the Gross-Pitaevskii equation gives a qualitatively correct
description of the dynamics.  It is also interesting to note that for
double-well condensates the Rabi regime is more difficult to achieve
since by reducing the hopping $J$ to achieve the condition in
Eq.~(\ref{Eq:condition1}), a single-band description becomes
inapplicable.  On the other hand, the Fock regime for double-well
condensates can be experimentally achieved, which has the
Mott Insulating ground state \cite{jaksch98, greiner02}.

\section{Semiclassical analysis of the rotor model}
\label{Sec:semiclassical}

\begin{figure*}
\includegraphics[width=7.in]{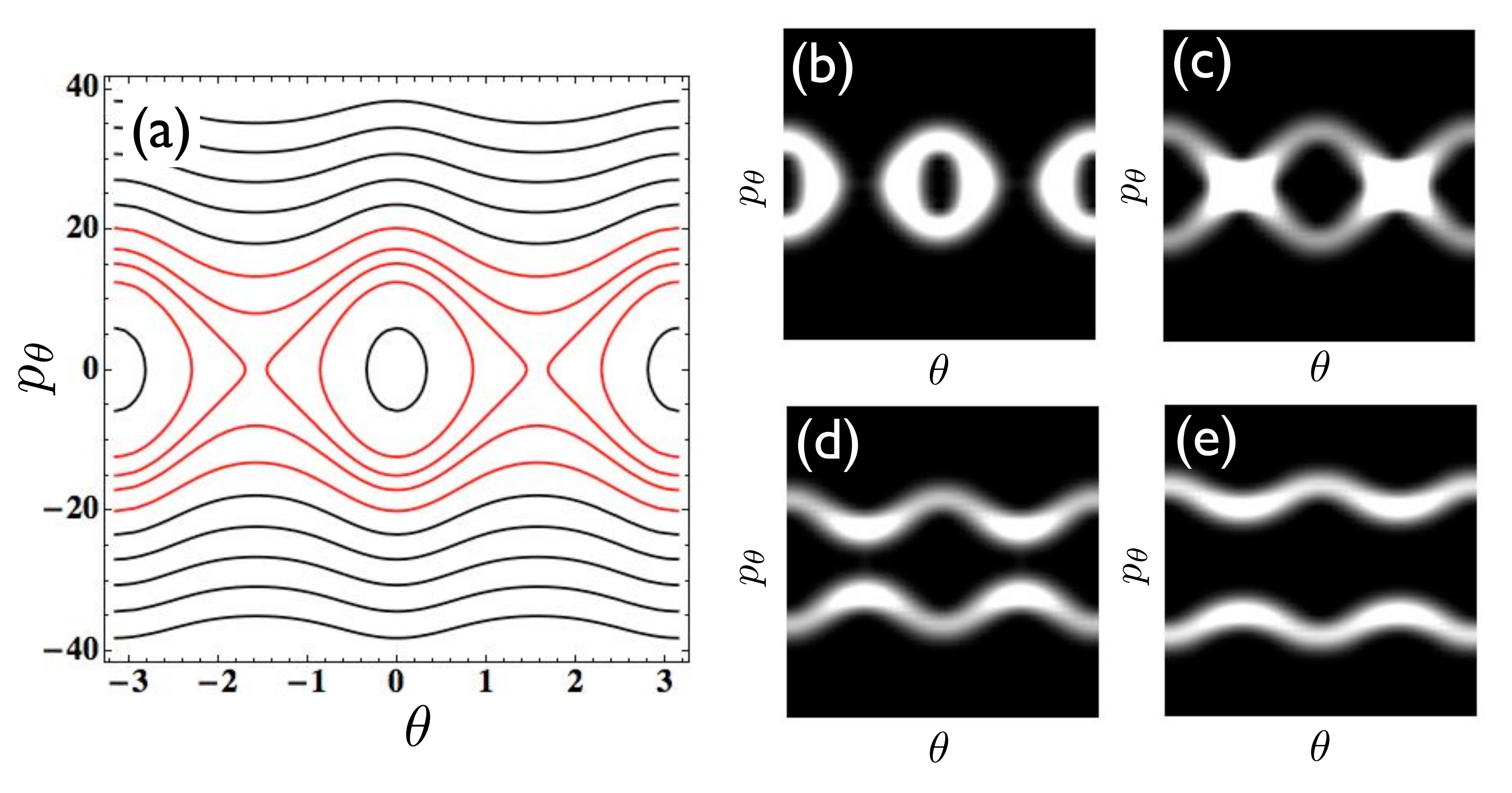}
\caption{Left: Equal-energy contours of the semiclassical energy given
in Eq.~(\ref{Eq:Ec}) for $p_\phi=0$. Right: Husimi distribution
functions H(\bz) for particular eigenstates of the rotor model
Eq.~(\ref{Eq:Hr}) for the parameters $q=g$, $N=10$, and $\kappa=1/10$.
Panels (b), (c), (d), (e) respectively correspond to the 2nd, 4th,
6th, and 8th excited states within the manifold of $m=0$ and even
$\ell$.  The classical equal-energy contours corresponding to the
energies of the states plotted on the right are shown in red.  The
same range of $p_\theta$ and $\theta$ is used for all plots.}
\label{Fig:montage}
\end{figure*}

In this Section we analyze the semiclassical limit of the rotor model
in Eq.~(\ref{Eq:Hr}).  We will use this to address recent
experimental results.  We will then show results from taking the
Husimi transform of the quantum eigenstates of the rotor model.  Such
a method has been shown to elucidate the semiclassical limit of the
Bose-Hubbard dimer model \cite{mahmud05}.

In recent experiments \cite{liu09a,liu09b} the dynamics of a $^{23}$Na
condensate, which has antiferromagnetic interactions $g>0$, was
investigated.  The initial condensate was prepared in a fully
polarized ferromagnetic state pointing in the $x$ direction after which
the condensate was allowed to freely
evolve.  The value of $\expect{F_x}^2$ was measured as a function of
time. Two distinct types of behavior were found, depending on the
external magnetic field which couples to the system through the
quadratic Zeeman shift $q$.  A separatrix between these two behaviors
occurs at a critical magnetic field $B_c$.  When $B<B_c$,
$\expect{F_x}^2$ showed oscillatory behavior, having
$\expect{F_x}^2>0$ at all times.  On the other hand, when $B>B_c$ it
was seen that $\expect{F_x}^2=0$ at periodic intervals during its
evolution.  An analysis of the behavior was provided in terms of the
classical Gross-Pitaevskii energy functional.  Taking into account the
conserved quantities (total particle number and spin moment in the
$z$-direction which was fixed to be $\expect{F_z}=0$), the
two-dimensional phase portrait of the energy functional was shown to
capture these two regimes.

We will now describe how the semiclassical limit of Eq.~(\ref{Eq:Hr}) 
gives an intuitive understanding of these results.  The corresponding
Lagrangian is
\begin{equation}
{\cal L} = \frac{1}{2} I \left(  \dot{\theta}^2  +
 \sin^2(\theta) \dot{\phi}^2 \right) - V(\theta)
\end{equation}
where $V(\theta)$ is given by Eq.~(\ref{Eq:V}).
This gives the canonical momenta $p_\theta=I\dot{\theta}$ and
$p_\phi= I \sin^2(\theta) \dot{\phi}$.  The corresponding classical
energy is
\begin{equation}
\label{Eq:Ec}
E=\frac{1}{2I} \left(  p_\theta^2 + \frac{p_\phi^2}{\sin^2(\theta)}
  \right) + V(\theta).
\end{equation}
The above equations describe the motion of a particle on a unit
sphere.  We will concentrate on the case where $p_\phi=0$.
The classical equal-energy contours of Eq.~(\ref{Eq:Ec}) are plotted
in the left panel of Fig.~\ref{Fig:montage}.  Two types of behavior are
seen.  The higher-energy states have motion where the
particle's trajectory explores both hemispheres but has either $p_\theta>0$
or $p_\theta<0$ at all times thus never having zero angular momentum.
This corresponds to the motion of the spin-one condensate for $B<B_c$.
Increasing the magnetic field will constrain the particle's trajectory to 
one hemisphere.  For this motion, it is seen that $p_\theta=0$ at periodic
intervals during the particle's trajectory.  This type of motion corresponds
to $B>B_c$ of the spin-one condensate experiment
\footnote{
It is  also worth pointing out that experiments in double-well potentials exhibiting
the self-trapping effect \cite{albiez05} can similarly be interpreted in 
terms of the classical phase space of Eq.~(\ref{Eq:rotor2}).  That is,
this phase space also exhibits a separatrix and trajectories with $p_\theta \ne 0$
at all times correspond to the self-trapped states}.

We will now move on to discuss the semiclassical properties manifest in the
quantum mechanical wave functions of Eq.~(\ref{Eq:Hr}) for appropriate
parameter regimes.  To do this, we will use a generalization of the
Husimi distribution function \cite{husimi40} to the case of the
sphere.  The Husimi distribution function has been successfully
applied to elucidate the pendulum structure manifest in the
wave functions of double-well condensates \cite{mahmud05} described by
Eq.~(\ref{Eq:doublewell}).

Conventionally, the Husimi distribution function of a particular wave
function $\ket{\psi}$ is defined as
\begin{equation} 
\label{Eq:Husimi}
H(\bz) = \frac{|\bra{\bz}\ketn{\psi}|^2}{\bra{\bz}\ketn{\bz}}
\end{equation} 
where $\ket{\bz}$ is a scaled coherent state.  For our
considerations, we thus need a generalization of the notion of a
coherent state to the unit sphere.  
Recent work on such a generalization is given
in Refs.~\cite{kowalski00, hall02}.  In  \cite{kowalski00} it was argued that it is most natural to define spherical
coherent states to be eigenstates of the ``annihilation'' operators
\begin{equation}
A_{\alpha}= e^{- \frac{1}{2} \kappa L^2}\Omega_\alpha e^{ \frac{1}{2} \kappa L^2}
\end{equation}
where $\kappa$ is a scaling parameter
\footnote{It is instructive to
compare this expression to $e^{-p^2/2} x e^{p^2/2}=x+ip$ (for
canonically conjugate operators $x$ and $p$) which has the
standard coherent states as eigenstates.}.  
Such eigenstates are
given by
\begin{equation}
\ket{\bz} = \sum_{\ell m} e^{- \kappa \ell(\ell+1)/2} \ket{Y_{\ell m}} Y^*_{\ell m} (\bz).
\end{equation}
In this equation $\bz$ is a three-component complex vector that satisfies
$\bz \cdot \bz=1$.  In terms of classical phase space variables 
(${\bf p} = p_\theta \hat{\theta} + p_\phi \hat{\phi}$  and $\bo$), 
$\bz$ can
be expressed as \cite{kowalski00}
\begin{equation}
\bz = \cosh(\kappa p) \bo + i \frac{1}{p} \sinh(\kappa p) {\bf p}.
\end{equation}
The value of the scaling parameter should be taken such that  $\kappa^2 \sim \frac{g}{qN^2}$ which
is the ratio of the prefactors of the kinetic and potential energies in Eq.~(\ref{Eq:Hr}).

We consider the case where $q=g$ and $N=10$ bosons which places the
system between the Josephson and Rabi regimes, and well away from the
Fock regime.  Density plots of the Husimi distribution function for
particular rotor eigenstates are shown in the right of
Fig.~\ref{Fig:montage}.  Since we concentrate on the case of $F_z=0$,
the Husimi distribution function will only depend on the pair of
variables $(\theta, p_\theta)$.  Red classical equal-energy contours
shown on the right of Fig.~\ref{Fig:montage} are drawn for energies
corresponding to these eigenstates.  One sees that the Husimi
distribution functions strongly resemble the semiclassical contours,
thus revealing the semiclassical behavior of the eigenstates.  Such
agreement occurs for all states in the Rabi and Josephson regimes, but
not for the Fock regime which has no semiclassical correspondence.

 \section{Extension to higher spins}
\label{Sec:ext}

We will now move on to discuss how to perform the rotor mapping
for larger spin systems.  We will focus on $F=2$ spinor
condensates because of their experimental availability as hyperfine
states of alkali atoms.  We will show that this system maps to a 
particle moving on a sphere in five-dimensional space.

Spin-two condensates have five spin components.  We take $a_{\alpha}$
for $\alpha=-2 \ldots 2$ to annihilate. a boson with $F_z=\alpha$.
In the single-mode regime, spin-two condensates are described by the
Hamiltonian \cite{ueda02}
\begin{equation}
\label{Eq:spintwo}
H= \frac{g_1}{2N} F^2 + \frac{g_2}{2N} \sA^\dagger \sA.
\end{equation}
Here,  ${\bf F}=a^\dagger_{\alpha} {\bf F}_{\alpha\beta}a_{\beta}$ is
the total spin operator where ${\bf F}_{\alpha \beta}$ are spin-two
matrices.  In the second term, $\sA=a_0 a_0 - 2 a_1 a_{-1} + 2 a_{1}
a_{-1}$ 
annihilates a singlet pair of bosons.  In terms of physical
quantities,
the coefficients $g_{1,2}$ are given by
\begin{align}
g_1&=\frac{4\pi
\hbar^2}{7m}n_0 (\bar{a}_4-\bar{a}_2)\\ 
g_2&=\frac{4\pi
\hbar^2}{m}n_0\left( \frac{1}{5} (\bar{a}_0-\bar{a}_4) - 
\frac{2}{7} (\bar{a}_2 -\bar{a}_4) \right)
\end{align}
where $\bar{a}_0$, $\bar{a}_2$, and $\bar{a}_4$ 
are the spin-two s-wave scattering lengths.
For simplicity we will neglect the effects of an external magnetic
field on this system.

We perform the following unitary transformation on the bosonic operators:
\begin{align}
b_{1} &=  a_0 \\
b_{2} &=  \frac{1}{i \sqrt{2}}(-a_{-1} - a_{1})\\
b_{3} &=  \frac{1}{\sqrt{2}}(a_{-1} - a_{1})\\
b_{4} &=  \frac{1}{i\sqrt{2}}(a_2 - a_{-2})\\
b_{5} &= \frac{1}{\sqrt{2}} (a_2 + a_{-2}).
\end{align}
These operators transform as a vector under SO(5) rotations generated
by $M_{\alpha \beta} = -i(b_\alpha^\dagger b_\beta - b_\beta^\dagger
b_\alpha)$.  In terms of these quantities, the singlet operator
is
\begin{equation}
\sA= {\bf b} \cdot {\bf b}
\end{equation}
while the spin operators are
\begin{align}
F_x &= \sqrt{3} M_{12} - M_{25} + M_{34} \\
F_y &= \sqrt{3} M_{13} + M_{24} + M_{35} \\
F_z &= M_{23} +2 M_{45}.
\end{align}

As before,  we can
parametrize an overcomplete set of states (but now using  the five-component,
real unit vector $\bo$) as 
\begin{equation}
\ket{\bo_N} = \frac{1}{\sqrt{N!}} 
\left( \sum_{\alpha=1}^5 \Omega_\alpha b_\alpha^\dagger \right) \ket{0}.
\end{equation}
The mapping proceeds along similar lines to that in 
Secs.~\ref{Sec:doublewell} and \ref{Sec:spinone}.  In particular,
one finds that
\begin{equation}
M_{\alpha \beta} \rightarrow - L_{\alpha \beta}
\end{equation}
where $L_{\alpha \beta}=-i(\Omega_\alpha \nabla_\beta -\Omega_\beta
\nabla_\alpha)$.
This can be used to find the rotor correspondence of the first term in
Eq.~(\ref{Eq:spintwo}).  

Next we find the rotor correspondence of
the second term in Eq.~(\ref{Eq:spintwo}).  We use the five-component
version of Eq.~(\ref{Eq:rule}) to find that 
\begin{align}
{\cal A}^\dagger {\cal A} \ket{\bo_N} = \left( \nabla^2 + N^2 + 3N\right) \ket{\bo_N}.
\end{align}
In this equation, $\nabla^2$ is the Laplacian on the five-dimensional
hypersphere, as described in Appendix \ref{A1}.  The
integration-by-parts here is straightforward.  One obtains
\begin{equation}
{\cal A}^\dagger {\cal A} \rightarrow \nabla^2 + N^2 + 3N.
\end{equation}
The resulting rotor model, which is already Hermitian, is thus
\begin{align}
\label{Eq:Hr2}
{\mathscr H} = &\frac{g_2}{2N} \nabla^2 +  \frac{g_1}{2N} \left(
(\sqrt{3} L_{12} - L_{25} + L_{34})^2+ \right. \\ 
&\qquad  \left.  (\sqrt{3} L_{13} + L_{24} +
L_{35} )^2
+ (L_{23} +2 L_{45})^2
\right)  \notag
\end{align}
where we have dropped a constant energy offset.  
This model has a particularly simple form in the limit
of $g_1=0$.  Here the system has an SO(5) symmetry, and the ground
state will be a condensate of singlet pairs of spin-two bosons.  
 
\section{Conclusion}
\label{Sec:conclusion}

In this work we have analyzed in detail rotor mappings of spinor
condensates in the single mode regime.  We have addressed some
subtleties related to the physical and unphysical eigenstates and
showed that the rotor mapping gives an exact treatment of the
spinor condensate.  Since the rotor model treats the mean field
as well as correlated phases on equal footing it offers new insights
into the problem.  We have established both the importance of the
rotor model in providing physical insight into the properties of
spinor condensates and its validity as a practical scheme for carrying
out calculations.

There are several interesting directions that can be pursued in future
work.  The Husimi distribution function has proven useful for
understanding the collapse and revival process of atoms in the
proximity of a superfluid-insulating phase transition
\cite{greiner02}.  Such a phase-space analysis of the collapse and
revival dynamics for the spinor system close to the Fock regime will
prove to be valuable.  We emphasize that for this regime the
semiclassical correspondence illustrated in Fig.~\ref{Fig:montage}
will not hold.

In Sec.~\ref{Sec:ext} we derived the rotor representation of the
spin-two system for a single site.  The mean-field phase diagram of
the spin-two condensates is known to have a degeneracy for nematic
states \cite{barnett06} which is lifted by quantum and thermal
fluctuations \cite{turner07,song07}.  A generalization
Eq.~(\ref{Eq:Hr2}) to include quadratic Zeeman field will prove useful
for studying this effect for smaller condensates where quantum effects
are more pronounced.  Finally, we note that low energy effective rotor
theories of spinor condensates (without magnetic fields) where
previously investigated in \cite{zhou01,demler02,imambekov03}.  It
will be interesting to investigate how the rotor mapping generalizes
to include spatial degrees of freedom.

\acknowledgements
It is our pleasure to acknowledge useful discussions with
A. Lamacraft, P. Lett, and W. Reinhardt.  This work was supported
by the NSF Joint Quantum Institute Physics Frontier Center.  We thank
the hospitality of the Kavli Institute of Theoretical Physics 
under the grant  NSF PHY05-51164 where part of this work was
completed.

 \appendix
 \section{Quantum mechanics on the hypersphere}
 \label{A1}
 
 In this Appendix, for convenience, we will tabulate the properties of
a particle constrained to the surface of a $d$-dimensional
hypersphere.  The position of the particle is given by $d$ coordinates
$\Omega_\alpha\ $ (for $\alpha=1,\ldots, d$) subject to the constraint
${\bo \cdot \bo}=1$.  The momentum operators are $\pi_\alpha = -i
\nabla_\alpha$ where $\nabla_\alpha$ is the gradient operator in the
$\alpha$-direction on the hypersphere (which can be expressed in terms
of $d-1$ angles and their derivatives).  Finally, the angular momentum
operators are $L_{\alpha \beta} = \Omega_\alpha \pi_\beta -
\Omega_\beta \pi_\alpha$.  Note that for $d=3$, the angular momentum
is conventionally written with a single subscript as $L_\alpha=
\frac{1}{2}\varepsilon_{\alpha \beta \gamma} L_{\beta \gamma}$.  The
position and angular momentum are Hermitian operators, while the
Hermitian conjugate of $\pi_\alpha$ is
 \begin{equation}
 \pi_\alpha^\dagger =  \pi_\alpha + i (d-1)\Omega_\alpha
 \end{equation}
 The following
 commutation relations are satisfied for the position and momentum operators:
 \begin{align}
 [\Omega_\alpha, \Omega_\beta] &= 0\\
 [\Omega_\alpha, \pi_\beta]  &= i(\delta_{\alpha \beta} - \Omega_\alpha \Omega_\beta)\\
 [\pi_\alpha, \pi_\beta] &=- i L_{\alpha \beta}.
 \end{align}
These can be seen to give the angular momentum operators the correct
commutation relations
which are $[L_{\alpha \beta}, L_{\gamma \delta}] =i
\delta_{\alpha \gamma}L_{\beta \delta}
+ i \delta_{\beta \delta} L_{\alpha \gamma} -i
\delta_{\alpha \delta} L_{\beta \gamma} - i \delta_{\beta
  \gamma} L_{\alpha \delta}$.
These operators satisfy the orthogonality relation ${\bf \Omega} \cdot {\bf \pi}=0$
(i.e. the momentum and position are orthogonal on the
hypersphere). It can also be verified that the total
angular momentum can be expressed as
\begin{align}
\frac{1}{2} \sum_{\alpha \beta}L_{\alpha \beta} L_{\beta \alpha} =
{\bf \pi} \cdot {\bf \pi} = - \nabla^2.
\end{align}

\section{Rotor model with general coupling}
\label{A2}

In this Appendix we consider spin-one Hamiltonians with more
general coupling to external fields.  In particular, we consider
Hamiltonians of the form
\begin{equation}
\label{Eq:Hgen} 
H = \frac{g}{2N} F^2 + b_{\alpha}^\dagger B_{\alpha \beta} b_\beta.
\end{equation} 
where $B$ is a Hermitian matrix.  One can see that this reduces to 
Eq.~(\ref{Eq:H}) for the special case $B_{\alpha \beta} = -q \delta_{\alpha z}
\delta_{\beta z}$.   In the following it is useful to write $B$ in
terms of its real and imaginary parts as $B=B' + i B''$.  Since $B$
is Hermitian, we have the requirement that $B'$ is symmetric while
$B''$ is antisymmetric.

Applying the rotor mapping as in Sec.~\ref{Sec:spinone} one
arrives at the non-Hermitian Hamiltonian
\begin{equation}
{\cal H}  = \frac{g}{2N} L^2  +  q(N+3)  B_{\alpha \beta} \Omega_\alpha \Omega_\beta
- B_{\alpha \beta} \Omega_\beta \nabla_\alpha
\end{equation}
which should be compared to Eq.~(\ref{Eq:Hrnh}).  To bring this
Hamiltonian to Hermitian form we apply the similarity transformation
${\mathscr H} = e^{-S} {\cal H} e^{S}$ where 
\begin{equation}
S=\Gamma_{\alpha \beta} \Omega_\alpha \Omega_\beta
\end{equation}
where $\Gamma$ is a matrix.  One can verify that by choosing  
$\Gamma = \frac{N}{2g} B'$, provided
$B'$ and $B''$ commute,  ${\mathscr H}$ becomes Hermitian.
In particular, for this value of $\Gamma$, 
\begin{equation}
{\mathscr H}= \frac{1}{2I} L^2 + V(\theta,\phi)
\end{equation} 
where  
\begin{align}
V(\theta,\phi) = &\left(N+\frac{3}{2} \right) \Omega^T B' \Omega
+\frac{1}{2}B''_{\alpha \beta} L_{\beta \alpha} \\ &+
\frac{N}{2g} \left(\Omega^T  B'^2 \Omega - (\Omega^T B' \Omega)^2\right).
\end{align}
One can check that this reduces to Eq.~(\ref{Eq:Hr}) in the appropriate
limit.

\section{The Bogoliubov spectrum of Eq.~(\ref{Eq:H})} 
\label{A3}

It is instructive to compute the low lying spectrum of the spinor
Hamiltonian Eq.~(\ref{Eq:H}) through the Bogoliubov method
\cite{bogoliubov47} and compare with the results from the exact rotor
mapping given in Sec.~\ref{Sec:spectra}.  
We expand about classical
mean-field state given by $\bar{a}_{1}=\bar{a}_{-1}=0$ and 
$\bar{a}_0=\sqrt{N}$, and write the bosonic operators as
$a_\alpha = \bar{a}_\alpha + \delta a_\alpha$.  The constraint of
fixed  particle number $N$ can be enforced up to quadratic order by
requiring 
\begin{align}
-\sqrt{N} (\delta a_0 + \delta a_0^\dagger) = \delta a_1^\dagger \delta
a_1 + \delta a_{-1}^\dagger \delta a_{-1}.
\end{align}
Dropping constant terms, Eq.~(\ref{Eq:H}) becomes up to quadratic
order 
\begin{align}
H = (g +q)(\delta a_1^\dagger \delta a_1 +\delta a_{-1}^\dagger \delta
a_{-1}) +  g (\delta a_1 \delta a_{-1} + {\rm h.c.}).
\end{align}
It is straightforward to diagonalize this by  a Bogoliubov
transformation.  The result is
\begin{equation}
H = \sqrt{q(2g+q)}( \alpha^\dagger \alpha + \beta^\dagger \beta)
\end{equation}
where $\alpha$ and $\beta$ are bosonic annihilation operators which is
consistent with \cite{cui08}.  The
spectrum of this hamiltonian can be seen to agree with the results
derived from the rotor model in Sec.~\ref{Sec:spectra} in the Rabi
regime, Eq.~(\ref{Eq:RabiE}), and Josephson regime,
Eq.~(\ref{Eq:JosephsonE}).  However, the results do not agree in the
Fock regime, Eq.~(\ref{Eq:FockE}), since the Bogoliubov treatment is
inapplicable for a fragmented condensate.

\section{Experimental realization of the singlet condensate}
\label{A4}

In this Appendix, we will discuss the experimental parameters
necessary to achieve the singlet condensate.  In the Josephson regime,
the ground state wave function of the rotor model is 
\begin{equation}
\psi(\theta)= \sqrt{\frac{2}{\pi\bar{\theta}^2}} e^{-\theta^2/\bar{\theta}^2}
\end{equation}
where $\bar{\theta}=\sqrt{\frac{2 g}{ qN^2}}$.  As can be verified from
(\ref{Eq:Josephson}), in the Josephson regime, $\bar{\theta} \ll 1$.
Decreasing the magnetic field and thereby decreasing $q$, one sees that
the width of the wave function increases.  When the width of the
wave function $\bar{\theta}$ approaches unity, the harmonic description
of the condensate fails and the Fock regime is entered
(\ref{Eq:Fock}).  As mentioned earlier, in the limiting case of $q=0$
the ground state is a condensate of singlet pairs of spin one bosons.

We thus ask what parameters are necessary for $\bar{\theta} \sim 1$.
Because this quantity scales inversely with the number of particles,
this state cannot be achieved in the thermodynamic limit.  We
therefore concentrate on systems with relatively small particle
number. 
For the $^{23}$Na system, the parameters $g$ and
$q$ appearing in the rotor model Eq.~(\ref{Eq:Hr}) are related to the
atomic density and external magnetic field  $B$ through \cite{liu09a}
\begin{align}
g &= \left( 1.59 \times 10^{-52} \rm{J m}^3  \right) n_0 \\
q &= \left( 1.84 \times 10^{-35} \rm{ J} / (\mu \rm{T})^2 \right)  B^2.
\end{align}

For a fixed particle number, increasing the density increases
$\bar{\theta}$.  We thus take $n_0=5 \times 10^{14}$cm$^{-3}$, which
is relatively large, but still small enough that three-body
losses are not important.  Then for small magnetic field $B=0.1 \mu
\rm{T}$ and $N=500$ particles, we have $\bar{\theta} = 1.9$ which is
outside of the Josephson regime.  If quenched from finite magnetic
field, such a system will exhibit quantum collapse and revival
oscillations \cite{barnett10}.


\begin{thebibliography}{10}%
\makeatletter
\providecommand \@ifxundefined [1]{%
 \ifx #1\undefined \expandafter \@firstoftwo
 \else \expandafter \@secondoftwo
\fi
}%
\providecommand \@ifnum [1]{%
 \ifnum #1\expandafter \@firstoftwo
 \else \expandafter \@secondoftwo
\fi
}%
\providecommand \enquote [1]{``#1''}%
\providecommand \bibnamefont  [1]{#1}%
\providecommand \bibfnamefont [1]{#1}%
\providecommand \citenamefont [1]{#1}%
\providecommand\href[0]{\@sanitize\@href}%
\providecommand\@href[1]{\endgroup\@@startlink{#1}\endgroup\@@href}%
\providecommand\@@href[1]{#1\@@endlink}%
\providecommand \@sanitize [0]{\begingroup\catcode`\&12\catcode`\#12\relax}%
\@ifxundefined \pdfoutput {\@firstoftwo}{%
 \@ifnum{\z@=\pdfoutput}{\@firstoftwo}{\@secondoftwo}%
}{%
 \providecommand\@@startlink[1]{\leavevmode}%
 \providecommand\@@endlink[0]{}%
}{%
 \providecommand\@@startlink[1]{%
  \leavevmode
  \pdfstartlink
   attr{/Border[0 0 1 ]/H/I/C[0 1 1]}%
   user{/Subtype/Link/A<</Type/Action/S/URI/URI(#1)>>}%
  \relax
 }%
 \providecommand\@@endlink[0]{\pdfendlink}%
}%
\providecommand \url  [0]{\begingroup\@sanitize \@url }%
\providecommand \@url [1]{\endgroup\@href {#1}{\urlprefix}}%
\providecommand \urlprefix [0]{URL }%
\providecommand \Eprint[0]{\href }%
\@ifxundefined \urlstyle {%
  \providecommand \doi [1]{doi:\discretionary{}{}{}#1}%
}{%
  \providecommand \doi [0]{doi:\discretionary{}{}{}\begingroup
  \urlstyle{rm}\Url }%
}%
\providecommand \doibase [0]{http://dx.doi.org/}%
\providecommand \Doi[1]{\href{\doibase#1}}%
\providecommand \bibAnnote [3]{%
  \BibitemShut{#1}%
  \begin{quotation}\noindent
    \textsc{Key:}\ #2\\\textsc{Annotation:}\ #3%
  \end{quotation}%
}%
\providecommand \bibAnnoteFile [2]{%
  \IfFileExists{#2}{\bibAnnote {#1} {#2} {\input{#2}}}{}%
}%
\providecommand \typeout [0]{\immediate \write \m@ne }%
\providecommand \selectlanguage [0]{\@gobble}%
\providecommand \bibinfo [0]{\@secondoftwo}%
\providecommand \bibfield [0]{\@secondoftwo}%
\providecommand \translation [1]{[#1]}%
\providecommand \BibitemOpen[0]{}%
\providecommand \bibitemStop [0]{}%
\providecommand \bibitemNoStop [0]{.\EOS\space}%
\providecommand \EOS [0]{\spacefactor3000\relax}%
\providecommand \BibitemShut [1]{\csname bibitem#1\endcsname}%
\bibitem{barnett10}%
  \BibitemOpen
  \bibfield{author}{%
  \bibinfo {author} {\bibfnamefont{R.}~\bibnamefont{Barnett}}, \bibinfo
  {author} {\bibfnamefont{J.~D.}\ \bibnamefont{Sau}},\ and\ \bibinfo {author}
  {\bibfnamefont{S.}~\bibnamefont{Das\;Sarma}},\ }%
  \bibfield{journal}{%
  \bibinfo {journal} {Phys. Rev. A}\ }%
  \textbf{\bibinfo {volume} {82}},\ \bibinfo {pages} {031602(R)} (\bibinfo
  {year} {2010})%
  \bibAnnoteFile{NoStop}{barnett10}%
\bibitem{josephson62}%
  \BibitemOpen
  \bibfield{author}{%
  \bibinfo {author} {\bibfnamefont{B.~D.}\ \bibnamefont{Josephson}},\ }%
  \bibfield{journal}{%
  \bibinfo {journal} {Phys. Lett.}\ }%
  \textbf{\bibinfo {volume} {1}},\ \bibinfo {pages} {251} (\bibinfo {year}
  {1962})%
  \bibAnnoteFile{NoStop}{josephson62}%
\bibitem{anderson67}%
  \BibitemOpen
  \bibinfo {note} {P.~W. in \emph{Progress in Low Temperature Physics vol V},
  ed C J Gorter (Amsterdam: North-Holland), p. 1-43}%
  \bibAnnoteFile{NoStop}{anderson67}%
\bibitem{tinkham04}%
  \BibitemOpen
  \bibfield{author}{%
  \bibinfo {author} {\bibfnamefont{M.}~\bibnamefont{Tinkham}},\ }%
  \emph{\bibinfo {title} {Introduction to Superconductivity}}\ (\bibinfo
  {publisher} {Dover},\ \bibinfo {year} {New York, 2004})%
  \bibAnnoteFile{NoStop}{tinkham04}%
\bibitem{sadler06}%
  \BibitemOpen
  \bibfield{author}{%
  \bibinfo {author} {\bibfnamefont{L.~E.}\ \bibnamefont{Sadler}}, \bibinfo
  {author} {\bibfnamefont{J.~M.}\ \bibnamefont{Higbie}}, \bibinfo {author}
  {\bibfnamefont{S.~R.}\ \bibnamefont{Leslie}}, \bibinfo {author}
  {\bibfnamefont{M.}~\bibnamefont{Vengalattore}},\ and\ \bibinfo {author}
  {\bibfnamefont{D.~M.}\ \bibnamefont{Stamper-Kurn}},\ }%
  \bibfield{journal}{%
  \bibinfo {journal} {Nature}\ }%
  \textbf{\bibinfo {volume} {443}},\ \bibinfo {pages} {312} (\bibinfo {year}
  {2006})%
  \bibAnnoteFile{NoStop}{sadler06}%
\bibitem{vengalatorre08}%
  \BibitemOpen
  \bibfield{author}{%
  \bibinfo {author} {\bibfnamefont{M.}~\bibnamefont{Vengalattore}}, \bibinfo
  {author} {\bibfnamefont{S.~R.}\ \bibnamefont{Leslie}}, \bibinfo {author}
  {\bibfnamefont{J.}~\bibnamefont{Guzman}},\ and\ \bibinfo {author}
  {\bibfnamefont{D.~M.}\ \bibnamefont{Stamper-Kurn}},\ }%
  \bibfield{journal}{%
  \bibinfo {journal} {Phys. Rev. Lett.}\ }%
  \textbf{\bibinfo {volume} {100}},\ \bibinfo {pages} {170403} (\bibinfo {year}
  {2008})%
  \bibAnnoteFile{NoStop}{vengalatorre08}%
\bibitem{klempt09}%
  \BibitemOpen
  \bibfield{author}{%
  \bibinfo {author} {\bibfnamefont{C.}~\bibnamefont{Klempt}}, \bibinfo {author}
  {\bibfnamefont{O.}~\bibnamefont{Topic}}, \bibinfo {author}
  {\bibfnamefont{G.}~\bibnamefont{Gebreyesus}}, \bibinfo {author}
  {\bibfnamefont{M.}~\bibnamefont{Scherer}}, \bibinfo {author}
  {\bibfnamefont{T.}~\bibnamefont{Henninger}}, \bibinfo {author}
  {\bibfnamefont{P.}~\bibnamefont{Hyllus}}, \bibinfo {author}
  {\bibfnamefont{W.}~\bibnamefont{Ertmer}}, \bibinfo {author}
  {\bibfnamefont{L.}~\bibnamefont{Santos}},\ and\ \bibinfo {author}
  {\bibfnamefont{J.~J.}\ \bibnamefont{Arlt}},\ }%
  \bibfield{journal}{%
  \bibinfo {journal} {Phys. Rev. Lett.}\ }%
  \textbf{\bibinfo {volume} {103}},\ \bibinfo {pages} {195302} (\bibinfo {year}
  {2009})%
  \bibAnnoteFile{NoStop}{klempt09}%
\bibitem{vengalattore10}%
  \BibitemOpen
  \bibfield{author}{%
  \bibinfo {author} {\bibfnamefont{M.}~\bibnamefont{Vengalattore}}, \bibinfo
  {author} {\bibfnamefont{J.}~\bibnamefont{Guzman}}, \bibinfo {author}
  {\bibfnamefont{S.~R.}\ \bibnamefont{Leslie}}, \bibinfo {author}
  {\bibfnamefont{F.}~\bibnamefont{Serwane}},\ and\ \bibinfo {author}
  {\bibfnamefont{D.~M.}\ \bibnamefont{Stamper-Kurn}},\ }%
  \bibfield{journal}{%
  \bibinfo {journal} {Phys. Rev. A}\ }%
  \textbf{\bibinfo {volume} {81}},\ \bibinfo {pages} {053612} (\bibinfo {year}
  {2010})%
  \bibAnnoteFile{NoStop}{vengalattore10}%
\bibitem{klempt10}%
  \BibitemOpen
  \bibfield{author}{%
  \bibinfo {author} {\bibfnamefont{C.}~\bibnamefont{Klempt}}, \bibinfo {author}
  {\bibfnamefont{O.}~\bibnamefont{Topic}}, \bibinfo {author}
  {\bibfnamefont{G.}~\bibnamefont{Gebreyesus}}, \bibinfo {author}
  {\bibfnamefont{M.}~\bibnamefont{Scherer}}, \bibinfo {author}
  {\bibfnamefont{T.}~\bibnamefont{Henninger}}, \bibinfo {author}
  {\bibfnamefont{P.}~\bibnamefont{Hyllus}}, \bibinfo {author}
  {\bibfnamefont{W.}~\bibnamefont{Ertmer}}, \bibinfo {author}
  {\bibfnamefont{L.}~\bibnamefont{Santos}},\ and\ \bibinfo {author}
  {\bibfnamefont{J.~J.}\ \bibnamefont{Arlt}},\ }%
  \bibfield{journal}{%
  \bibinfo {journal} {Phys. Rev. Lett.}\ }%
  \textbf{\bibinfo {volume} {104}},\ \bibinfo {pages} {195303} (\bibinfo {year}
  {2010})%
  \bibAnnoteFile{NoStop}{klempt10}%
\bibitem{kronjager10}%
  \BibitemOpen
  \bibfield{author}{%
  \bibinfo {author} {\bibfnamefont{J.}~\bibnamefont{Kronjager}}, \bibinfo
  {author} {\bibfnamefont{C.}~\bibnamefont{Becker}}, \bibinfo {author}
  {\bibfnamefont{P.}~\bibnamefont{Soltan-Panahi}}, \bibinfo {author}
  {\bibfnamefont{K.}~\bibnamefont{Bongs}},\ and\ \bibinfo {author}
  {\bibfnamefont{K.}~\bibnamefont{Sengstock}},\ }%
  \bibfield{journal}{%
  \bibinfo {journal} {Phys. Rev. Lett.}\ }%
  \textbf{\bibinfo {volume} {105}},\ \bibinfo {pages} {090402} (\bibinfo {year}
  {2010})%
  \bibAnnoteFile{NoStop}{kronjager10}%
\bibitem{stenger98}%
  \BibitemOpen
  \bibfield{author}{%
  \bibinfo {author} {\bibfnamefont{J.}~\bibnamefont{Stenger}}, \bibinfo
  {author} {\bibfnamefont{S.}~\bibnamefont{Inouye}}, \bibinfo {author}
  {\bibfnamefont{D.~M.}\ \bibnamefont{Stamper-Kurn}}, \bibinfo {author}
  {\bibfnamefont{H.~J.}\ \bibnamefont{Miesner}}, \bibinfo {author}
  {\bibfnamefont{A.~P.}\ \bibnamefont{Chikkatur}},\ and\ \bibinfo {author}
  {\bibfnamefont{W.}~\bibnamefont{Ketterle}},\ }%
  \bibfield{journal}{%
  \bibinfo {journal} {Nature}\ }%
  \textbf{\bibinfo {volume} {396}},\ \bibinfo {pages} {345} (\bibinfo {year}
  {1998})%
  \bibAnnoteFile{NoStop}{stenger98}%
\bibitem{chang04}%
  \BibitemOpen
  \bibfield{author}{%
  \bibinfo {author} {\bibfnamefont{M.~S.}\ \bibnamefont{Chang}}, \bibinfo
  {author} {\bibfnamefont{C.~D.}\ \bibnamefont{Hamley}}, \bibinfo {author}
  {\bibfnamefont{M.~D.}\ \bibnamefont{Barrett}}, \bibinfo {author}
  {\bibfnamefont{J.~A.}\ \bibnamefont{Sauer}}, \bibinfo {author}
  {\bibfnamefont{K.~M.}\ \bibnamefont{Fortier}}, \bibinfo {author}
  {\bibfnamefont{W.}~\bibnamefont{Zhang}}, \bibinfo {author}
  {\bibfnamefont{L.}~\bibnamefont{You}},\ and\ \bibinfo {author}
  {\bibfnamefont{M.~S.}\ \bibnamefont{Chapman}},\ }%
  \bibfield{journal}{%
  \bibinfo {journal} {Phys. Rev. Lett.}\ }%
  \textbf{\bibinfo {volume} {92}},\ \bibinfo {pages} {140403} (\bibinfo {year}
  {2004})%
  \bibAnnoteFile{NoStop}{chang04}%
\bibitem{schmaljohann04}%
  \BibitemOpen
  \bibfield{author}{%
  \bibinfo {author} {\bibfnamefont{H.}~\bibnamefont{Schmaljohann}}, \bibinfo
  {author} {\bibfnamefont{M.}~\bibnamefont{Erhard}}, \bibinfo {author}
  {\bibfnamefont{J.}~\bibnamefont{Kronjager}}, \bibinfo {author}
  {\bibfnamefont{M.}~\bibnamefont{Kottke}}, \bibinfo {author}
  {\bibfnamefont{S.}~\bibnamefont{van Staa}}, \bibinfo {author}
  {\bibfnamefont{L.}~\bibnamefont{Cacciapuoti}}, \bibinfo {author}
  {\bibfnamefont{J.~J.}\ \bibnamefont{Arlt}}, \bibinfo {author}
  {\bibfnamefont{K.}~\bibnamefont{Bongs}},\ and\ \bibinfo {author}
  {\bibfnamefont{K.}~\bibnamefont{Sengstock}},\ }%
  \bibfield{journal}{%
  \bibinfo {journal} {Phys. Rev. Lett.}\ }%
  \textbf{\bibinfo {volume} {92}},\ \bibinfo {pages} {040402} (\bibinfo {year}
  {2004})%
  \bibAnnoteFile{NoStop}{schmaljohann04}%
\bibitem{chang05}%
  \BibitemOpen
  \bibfield{author}{%
  \bibinfo {author} {\bibfnamefont{M.~S.}\ \bibnamefont{Chang}}, \bibinfo
  {author} {\bibfnamefont{Q.~S.}\ \bibnamefont{Qin}}, \bibinfo {author}
  {\bibfnamefont{W.~X.}\ \bibnamefont{Zhang}}, \bibinfo {author}
  {\bibfnamefont{L.}~\bibnamefont{You}},\ and\ \bibinfo {author}
  {\bibfnamefont{M.~S.}\ \bibnamefont{Chapman}},\ }%
  \bibfield{journal}{%
  \bibinfo {journal} {Nature Phys.}\ }%
  \textbf{\bibinfo {volume} {1}},\ \bibinfo {pages} {111} (\bibinfo {year}
  {2005})%
  \bibAnnoteFile{NoStop}{chang05}%
\bibitem{widera05}%
  \BibitemOpen
  \bibfield{author}{%
  \bibinfo {author} {\bibfnamefont{A.}~\bibnamefont{Widera}}, \bibinfo {author}
  {\bibfnamefont{F.}~\bibnamefont{Gerbier}}, \bibinfo {author}
  {\bibfnamefont{S.}~\bibnamefont{Folling}}, \bibinfo {author}
  {\bibfnamefont{T.}~\bibnamefont{Gericke}}, \bibinfo {author}
  {\bibfnamefont{O.}~\bibnamefont{Mandel}},\ and\ \bibinfo {author}
  {\bibfnamefont{I.}~\bibnamefont{Bloch}},\ }%
  \bibfield{journal}{%
  \bibinfo {journal} {Phys. Rev. Lett.}\ }%
  \textbf{\bibinfo {volume} {95}},\ \bibinfo {pages} {190405} (\bibinfo {year}
  {2005})%
  \bibAnnoteFile{NoStop}{widera05}%
\bibitem{mur-petit06}%
  \BibitemOpen
  \bibfield{author}{%
  \bibinfo {author} {\bibfnamefont{J.}~\bibnamefont{Mur-Petit}}, \bibinfo
  {author} {\bibfnamefont{M.}~\bibnamefont{Guilleumas}}, \bibinfo {author}
  {\bibfnamefont{A.}~\bibnamefont{Polls}}, \bibinfo {author}
  {\bibfnamefont{A.}~\bibnamefont{Sanpera}}, \bibinfo {author}
  {\bibfnamefont{M.}~\bibnamefont{Lewenstein}}, \bibinfo {author}
  {\bibfnamefont{K.}~\bibnamefont{Bongs}},\ and\ \bibinfo {author}
  {\bibfnamefont{K.}~\bibnamefont{Sengstock}},\ }%
  \bibfield{journal}{%
  \bibinfo {journal} {Phys. Rev. A}\ }%
  \textbf{\bibinfo {volume} {73}},\ \bibinfo {pages} {013629} (\bibinfo {year}
  {2006})%
  \bibAnnoteFile{NoStop}{mur-petit06}%
\bibitem{gerbier06}%
  \BibitemOpen
  \bibfield{author}{%
  \bibinfo {author} {\bibfnamefont{F.}~\bibnamefont{Gerbier}}, \bibinfo
  {author} {\bibfnamefont{A.}~\bibnamefont{Widera}}, \bibinfo {author}
  {\bibfnamefont{S.}~\bibnamefont{Folling}}, \bibinfo {author}
  {\bibfnamefont{O.}~\bibnamefont{Mandel}},\ and\ \bibinfo {author}
  {\bibfnamefont{I.}~\bibnamefont{Bloch}},\ }%
  \bibfield{journal}{%
  \bibinfo {journal} {Phys. Rev. A}\ }%
  \textbf{\bibinfo {volume} {73}},\ \bibinfo {pages} {041602} (\bibinfo {year}
  {2006})%
  \bibAnnoteFile{NoStop}{gerbier06}%
\bibitem{liu09a}%
  \BibitemOpen
  \bibfield{author}{%
  \bibinfo {author} {\bibfnamefont{Y.}~\bibnamefont{Liu}}, \bibinfo {author}
  {\bibfnamefont{S.}~\bibnamefont{Jung}}, \bibinfo {author}
  {\bibfnamefont{S.~E.}\ \bibnamefont{Maxwell}}, \bibinfo {author}
  {\bibfnamefont{L.~D.}\ \bibnamefont{Turner}}, \bibinfo {author}
  {\bibfnamefont{E.}~\bibnamefont{Tiesinga}},\ and\ \bibinfo {author}
  {\bibfnamefont{P.~D.}\ \bibnamefont{Lett}},\ }%
  \bibfield{journal}{%
  \bibinfo {journal} {Phys. Rev. Lett.}\ }%
  \textbf{\bibinfo {volume} {102}},\ \bibinfo {pages} {125301} (\bibinfo {year}
  {2009})%
  \bibAnnoteFile{NoStop}{liu09a}%
\bibitem{liu09b}%
  \BibitemOpen
  \bibfield{author}{%
  \bibinfo {author} {\bibfnamefont{Y.}~\bibnamefont{Liu}}, \bibinfo {author}
  {\bibfnamefont{E.}~\bibnamefont{Gomez}}, \bibinfo {author}
  {\bibfnamefont{S.~E.}\ \bibnamefont{Maxwell}}, \bibinfo {author}
  {\bibfnamefont{L.~D.}\ \bibnamefont{Turner}}, \bibinfo {author}
  {\bibfnamefont{E.}~\bibnamefont{Tiesinga}},\ and\ \bibinfo {author}
  {\bibfnamefont{P.~D.}\ \bibnamefont{Lett}},\ }%
  \bibfield{journal}{%
  \bibinfo {journal} {Phys. Rev. Lett.}\ }%
  \textbf{\bibinfo {volume} {102}},\ \bibinfo {pages} {225301} (\bibinfo {year}
  {2009})%
  \bibAnnoteFile{NoStop}{liu09b}%
\bibitem{law98}%
  \BibitemOpen
  \bibfield{author}{%
  \bibinfo {author} {\bibfnamefont{C.~K.}\ \bibnamefont{Law}}, \bibinfo
  {author} {\bibfnamefont{H.}~\bibnamefont{Pu}},\ and\ \bibinfo {author}
  {\bibfnamefont{N.~P.}\ \bibnamefont{Bigelow}},\ }%
  \bibfield{journal}{%
  \bibinfo {journal} {Phys. Rev. Lett.}\ }%
  \textbf{\bibinfo {volume} {81}},\ \bibinfo {pages} {5257} (\bibinfo {year}
  {1998})%
  \bibAnnoteFile{NoStop}{law98}%
\bibitem{ho00}%
  \BibitemOpen
  \bibfield{author}{%
  \bibinfo {author} {\bibfnamefont{T.~L.}\ \bibnamefont{Ho}}\ and\ \bibinfo
  {author} {\bibfnamefont{S.~K.}\ \bibnamefont{Yip}},\ }%
  \bibfield{journal}{%
  \bibinfo {journal} {Phys. Rev. Lett.}\ }%
  \textbf{\bibinfo {volume} {84}},\ \bibinfo {pages} {4031} (\bibinfo {year}
  {2000})%
  \bibAnnoteFile{NoStop}{ho00}%
\bibitem{mueller06}%
  \BibitemOpen
  \bibfield{author}{%
  \bibinfo {author} {\bibfnamefont{E.~J.}\ \bibnamefont{Mueller}}, \bibinfo
  {author} {\bibfnamefont{T.-L.}\ \bibnamefont{Ho}}, \bibinfo {author}
  {\bibfnamefont{M.}~\bibnamefont{Ueda}},\ and\ \bibinfo {author}
  {\bibfnamefont{G.}~\bibnamefont{Baym}},\ }%
  \bibfield{journal}{%
  \bibinfo {journal} {Phys. Rev. A}\ }%
  \textbf{\bibinfo {volume} {74}},\ \bibinfo {pages} {033612} (\bibinfo {year}
  {2006})%
  \bibAnnoteFile{NoStop}{mueller06}%
\bibitem{ho98}%
  \BibitemOpen
  \bibfield{author}{%
  \bibinfo {author} {\bibfnamefont{T.-L.}\ \bibnamefont{Ho}},\ }%
  \bibfield{journal}{%
  \bibinfo {journal} {Phys. Rev. Lett.}\ }%
  \textbf{\bibinfo {volume} {81}},\ \bibinfo {pages} {742} (\bibinfo {year}
  {1998})%
  \bibAnnoteFile{NoStop}{ho98}%
\bibitem{ohmi98}%
  \BibitemOpen
  \bibfield{author}{%
  \bibinfo {author} {\bibfnamefont{T.}~\bibnamefont{Ohmi}}\ and\ \bibinfo
  {author} {\bibfnamefont{K.}~\bibnamefont{Machida}},\ }%
  \bibfield{journal}{%
  \bibinfo {journal} {J. Phys. Soc. Japan}\ }%
  \textbf{\bibinfo {volume} {67}},\ \bibinfo {pages} {1822} (\bibinfo {year}
  {1998})%
  \bibAnnoteFile{NoStop}{ohmi98}%
\bibitem{barnett09}%
  \BibitemOpen
  \bibfield{author}{%
  \bibinfo {author} {\bibfnamefont{R.}~\bibnamefont{Barnett}}, \bibinfo
  {author} {\bibfnamefont{D.}~\bibnamefont{Podolsky}},\ and\ \bibinfo {author}
  {\bibfnamefont{G.}~\bibnamefont{Refael}},\ }%
  \bibfield{journal}{%
  \bibinfo {journal} {Phys. Rev. B}\ }%
  \textbf{\bibinfo {volume} {80}},\ \bibinfo {pages} {024420} (\bibinfo {year}
  {2009})%
  \bibAnnoteFile{NoStop}{barnett09}%
\bibitem{koashi00}%
  \BibitemOpen
  \bibfield{author}{%
  \bibinfo {author} {\bibfnamefont{M.}~\bibnamefont{Koashi}}\ and\ \bibinfo
  {author} {\bibfnamefont{M.}~\bibnamefont{Ueda}},\ }%
  \bibfield{journal}{%
  \bibinfo {journal} {Phys. Rev. Lett.}\ }%
  \textbf{\bibinfo {volume} {84}},\ \bibinfo {pages} {1066} (\bibinfo {year}
  {2000})%
  \bibAnnoteFile{NoStop}{koashi00}%
\bibitem{ashhab02}%
  \BibitemOpen
  \bibfield{author}{%
  \bibinfo {author} {\bibfnamefont{S.}~\bibnamefont{Ashhab}}\ and\ \bibinfo
  {author} {\bibfnamefont{A.~J.}\ \bibnamefont{Leggett}},\ }%
  \bibfield{journal}{%
  \bibinfo {journal} {Phys. Rev. A}\ }%
  \textbf{\bibinfo {volume} {65}},\ \bibinfo {pages} {023604} (\bibinfo {year}
  {2002})%
  \bibAnnoteFile{NoStop}{ashhab02}%
\bibitem{diener06}%
  \BibitemOpen
  \bibfield{author}{%
  \bibinfo {author} {\bibfnamefont{R.}~\bibnamefont{Diener}}\ and\ \bibinfo
  {author} {\bibfnamefont{T.-L.}\ \bibnamefont{Ho}},\ }%
  \bibinfo {note} {arXiv:cond-mat/0608732}%
  \bibAnnoteFile{NoStop}{diener06}%
\bibitem{chang07}%
  \BibitemOpen
  \bibfield{author}{%
  \bibinfo {author} {\bibfnamefont{L.}~\bibnamefont{Chang}}, \bibinfo {author}
  {\bibfnamefont{Q.}~\bibnamefont{Zhai}}, \bibinfo {author}
  {\bibfnamefont{R.}~\bibnamefont{Lu}},\ and\ \bibinfo {author}
  {\bibfnamefont{L.}~\bibnamefont{You}},\ }%
  \bibfield{journal}{%
  \bibinfo {journal} {Phys. Rev. Lett.}\ }%
  \textbf{\bibinfo {volume} {99}},\ \bibinfo {pages} {080402} (\bibinfo {year}
  {2007})%
  \bibAnnoteFile{NoStop}{chang07}%
\bibitem{cui08}%
  \BibitemOpen
  \bibfield{author}{%
  \bibinfo {author} {\bibfnamefont{X.}~\bibnamefont{Cui}}, \bibinfo {author}
  {\bibfnamefont{Y.}~\bibnamefont{Wang}},\ and\ \bibinfo {author}
  {\bibfnamefont{F.}~\bibnamefont{Zhou}},\ }%
  \bibfield{journal}{%
  \bibinfo {journal} {Phys. Rev. A}\ }%
  \textbf{\bibinfo {volume} {78}},\ \bibinfo {pages} {050701(R)} (\bibinfo
  {year} {2008})%
  \bibAnnoteFile{NoStop}{cui08}%
\bibitem{zhai09}%
  \BibitemOpen
  \bibfield{author}{%
  \bibinfo {author} {\bibfnamefont{Q.}~\bibnamefont{Zhai}}, \bibinfo {author}
  {\bibfnamefont{L.}~\bibnamefont{Chang}}, \bibinfo {author}
  {\bibfnamefont{R.}~\bibnamefont{Lu}},\ and\ \bibinfo {author}
  {\bibfnamefont{L.}~\bibnamefont{You}},\ }%
  \bibfield{journal}{%
  \bibinfo {journal} {Phys. Rev. A}\ }%
  \textbf{\bibinfo {volume} {79}},\ \bibinfo {pages} {043608} (\bibinfo {year}
  {2009})%
  \bibAnnoteFile{NoStop}{zhai09}%
\bibitem{anglin01}%
  \BibitemOpen
  \bibfield{author}{%
  \bibinfo {author} {\bibfnamefont{J.~R.}\ \bibnamefont{Anglin}}, \bibinfo
  {author} {\bibfnamefont{P.}~\bibnamefont{Drummond}},\ and\ \bibinfo {author}
  {\bibfnamefont{A.}~\bibnamefont{Smerzi}},\ }%
  \bibfield{journal}{%
  \bibinfo {journal} {Phys. Rev. A}\ }%
  \textbf{\bibinfo {volume} {64}},\ \bibinfo {pages} {063605} (\bibinfo {year}
  {2001})%
  \bibAnnoteFile{NoStop}{anglin01}%
\bibitem{husimi40}%
  \BibitemOpen
  \bibfield{author}{%
  \bibinfo {author} {\bibfnamefont{K.}~\bibnamefont{Husimi}},\ }%
  \bibfield{journal}{%
  \bibinfo {journal} {Proc. Physico-Math Soc. Japan}\ }%
  \textbf{\bibinfo {volume} {22}},\ \bibinfo {pages} {264} (\bibinfo {year}
  {1940})%
  \bibAnnoteFile{NoStop}{husimi40}%
\bibitem{mahmud05}%
  \BibitemOpen
  \bibfield{author}{%
  \bibinfo {author} {\bibfnamefont{K.~W.}\ \bibnamefont{Mahmud}}, \bibinfo
  {author} {\bibfnamefont{H.}~\bibnamefont{Perry}},\ and\ \bibinfo {author}
  {\bibfnamefont{W.~P.}\ \bibnamefont{Reinhardt1}},\ }%
  \bibfield{journal}{%
  \bibinfo {journal} {Phys. Rev. A}\ }%
  \textbf{\bibinfo {volume} {71}},\ \bibinfo {pages} {023615} (\bibinfo {year}
  {2005})%
  \bibAnnoteFile{NoStop}{mahmud05}%
\bibitem{milburn97}%
  \BibitemOpen
  \bibfield{author}{%
  \bibinfo {author} {\bibfnamefont{G.~J.}\ \bibnamefont{Milburn}}, \bibinfo
  {author} {\bibfnamefont{J.}~\bibnamefont{Corney}}, \bibinfo {author}
  {\bibfnamefont{E.~M.}\ \bibnamefont{Wright}},\ and\ \bibinfo {author}
  {\bibfnamefont{D.~F.}\ \bibnamefont{Walls}},\ }%
  \bibfield{journal}{%
  \bibinfo {journal} {Phys. Rev. A}\ }%
  \textbf{\bibinfo {volume} {55}},\ \bibinfo {pages} {4318} (\bibinfo {year}
  {1997})%
  \bibAnnoteFile{NoStop}{milburn97}%
\bibitem{smerzi97}%
  \BibitemOpen
  \bibfield{author}{%
  \bibinfo {author} {\bibfnamefont{A.}~\bibnamefont{Smerzi}}, \bibinfo {author}
  {\bibfnamefont{S.}~\bibnamefont{Fantoni}}, \bibinfo {author}
  {\bibfnamefont{S.}~\bibnamefont{Giovanazzi}},\ and\ \bibinfo {author}
  {\bibfnamefont{S.~R.}\ \bibnamefont{Shenoy}},\ }%
  \bibfield{journal}{%
  \bibinfo {journal} {Phys. Rev. Lett.}\ }%
  \textbf{\bibinfo {volume} {79}},\ \bibinfo {pages} {4950} (\bibinfo {year}
  {1997})%
  \bibAnnoteFile{NoStop}{smerzi97}%
\bibitem{leggett01}%
  \BibitemOpen
  \bibfield{author}{%
  \bibinfo {author} {\bibfnamefont{A.~J.}\ \bibnamefont{Leggett}},\ }%
  \bibfield{journal}{%
  \bibinfo {journal} {Rev. Modern Phys.}\ }%
  \textbf{\bibinfo {volume} {73}},\ \bibinfo {pages} {307} (\bibinfo {year}
  {2001})%
  \bibAnnoteFile{NoStop}{leggett01}%
\bibitem{albiez05}%
  \BibitemOpen
  \bibfield{author}{%
  \bibinfo {author} {\bibfnamefont{M.}~\bibnamefont{Albiez}}, \bibinfo {author}
  {\bibfnamefont{R.}~\bibnamefont{Gati}}, \bibinfo {author}
  {\bibfnamefont{J.}~\bibnamefont{F\"olling}}, \bibinfo {author}
  {\bibfnamefont{S.}~\bibnamefont{Hunsmann}}, \bibinfo {author}
  {\bibfnamefont{M.}~\bibnamefont{Cristiani}},\ and\ \bibinfo {author}
  {\bibfnamefont{M.~K.}\ \bibnamefont{Oberthaler}},\ }%
  \bibfield{journal}{%
  \bibinfo {journal} {Phys. Rev. Lett.}\ }%
  \textbf{\bibinfo {volume} {95}},\ \bibinfo {pages} {010402} (\bibinfo {year}
  {2005})%
  \bibAnnoteFile{NoStop}{albiez05}%
\bibitem{bargmann61}%
  \BibitemOpen
  \bibfield{author}{%
  \bibinfo {author} {\bibfnamefont{V.}~\bibnamefont{Bargmann}},\ }%
  \bibfield{journal}{%
  \bibinfo {journal} {Comm. Pure Appl. Mathematics}\ }%
  \textbf{\bibinfo {volume} {14}},\ \bibinfo {pages} {187} (\bibinfo {year}
  {1961})%
  \bibAnnoteFile{NoStop}{bargmann61}%
\bibitem{Note1}%
  \BibitemOpen
  \bibinfo {note} {This can be shown by assuming two solutions to Eq.~(\ref
  {Eq:Hr}) with the same energy. It can be shown that the resulting Wronskian
  vanishes and thus the two solutions are equal to each other up to a
  multiplicative constant}%
  \bibAnnoteFile{NoStop}{Note1}%
\bibitem{leggett98}%
  \BibitemOpen
  \bibinfo {note} {A. J. Leggett in \emph{Proceedings of the 16th International
  Conference on Atomic Physics}, Windsor, Ontario, Canada, 1998, edited by W.
  E. Baylis and G. F. Drake (AIP, Woodbury, New York)}%
  \bibAnnoteFile{NoStop}{leggett98}%
\bibitem{jaksch98}%
  \BibitemOpen
  \bibfield{author}{%
  \bibinfo {author} {\bibfnamefont{D.}~\bibnamefont{Jaksch}}, \bibinfo {author}
  {\bibfnamefont{C.}~\bibnamefont{Bruder}}, \bibinfo {author}
  {\bibfnamefont{J.~I.}\ \bibnamefont{Cirac}}, \bibinfo {author}
  {\bibfnamefont{C.~W.}\ \bibnamefont{Gardiner}},\ and\ \bibinfo {author}
  {\bibfnamefont{P.}~\bibnamefont{Zoller}},\ }%
  \bibfield{journal}{%
  \bibinfo {journal} {Phys. Rev. Lett.}\ }%
  \textbf{\bibinfo {volume} {81}},\ \bibinfo {pages} {3108} (\bibinfo {year}
  {1998})%
  \bibAnnoteFile{NoStop}{jaksch98}%
\bibitem{greiner02}%
  \BibitemOpen
  \bibfield{author}{%
  \bibinfo {author} {\bibfnamefont{M.}~\bibnamefont{Greiner}}, \bibinfo
  {author} {\bibfnamefont{O.}~\bibnamefont{Mandel}}, \bibinfo {author}
  {\bibfnamefont{T.}~\bibnamefont{Esslinger}}, \bibinfo {author}
  {\bibfnamefont{T.~W.}\ \bibnamefont{Hansch}},\ and\ \bibinfo {author}
  {\bibfnamefont{I.}~\bibnamefont{Bloch}},\ }%
  \bibfield{journal}{%
  \bibinfo {journal} {Nature}\ }%
  \textbf{\bibinfo {volume} {415}},\ \bibinfo {pages} {39} (\bibinfo {year}
  {2002})%
  \bibAnnoteFile{NoStop}{greiner02}%
\bibitem{Note2}%
  \BibitemOpen
  \bibinfo {note} {It is also worth pointing out that experiments in
  double-well potentials exhibiting the self-trapping effect \cite {albiez05}
  can similarly be interpreted in terms of the classical phase space of
  Eq.~(\ref {Eq:rotor2}). That is, this phase space also exhibits a separatrix
  and trajectories with $p_\theta \not =0$ at all times correspond to the
  self-trapped states}%
  \bibAnnoteFile{NoStop}{Note2}%
\bibitem{kowalski00}%
  \BibitemOpen
  \bibfield{author}{%
  \bibinfo {author} {\bibfnamefont{K.}~\bibnamefont{Kowalski}}\ and\ \bibinfo
  {author} {\bibfnamefont{J.}~\bibnamefont{Rembieli\'{n}ski}},\ }%
  \bibfield{journal}{%
  \bibinfo {journal} {J. Phys. A: Math. Gen.}\ }%
  \textbf{\bibinfo {volume} {33}},\ \bibinfo {pages} {6035} (\bibinfo {year}
  {2000})%
  \bibAnnoteFile{NoStop}{kowalski00}%
\bibitem{hall02}%
  \BibitemOpen
  \bibfield{author}{%
  \bibinfo {author} {\bibfnamefont{B.}~\bibnamefont{Hall}}\ and\ \bibinfo
  {author} {\bibfnamefont{J.~J.}\ \bibnamefont{Mitchell}},\ }%
  \bibfield{journal}{%
  \bibinfo {journal} {J. Math. Phys.}\ }%
  \textbf{\bibinfo {volume} {43}},\ \bibinfo {pages} {1211} (\bibinfo {year}
  {2002})%
  \bibAnnoteFile{NoStop}{hall02}%
\bibitem{Note3}%
  \BibitemOpen
  \bibinfo {note} {It is instructive to compare this expression to $e^{-p^2/2}
  x e^{p^2/2}=x+ip$ (for canonically conjugate operators $x$ and $p$) which has
  the standard coherent states as eigenstates.}%
  \bibAnnoteFile{Stop}{Note3}%
\bibitem{ueda02}%
  \BibitemOpen
  \bibfield{author}{%
  \bibinfo {author} {\bibfnamefont{M.}~\bibnamefont{Ueda}}\ and\ \bibinfo
  {author} {\bibfnamefont{M.}~\bibnamefont{Koashi}},\ }%
  \bibfield{journal}{%
  \bibinfo {journal} {Phys. Rev. A}\ }%
  \textbf{\bibinfo {volume} {65}},\ \bibinfo {pages} {063602} (\bibinfo {year}
  {2002})%
  \bibAnnoteFile{NoStop}{ueda02}%
\bibitem{barnett06}%
  \BibitemOpen
  \bibfield{author}{%
  \bibinfo {author} {\bibfnamefont{R.}~\bibnamefont{Barnett}}, \bibinfo
  {author} {\bibfnamefont{A.}~\bibnamefont{Turner}},\ and\ \bibinfo {author}
  {\bibfnamefont{E.}~\bibnamefont{Demler}},\ }%
  \bibfield{journal}{%
  \bibinfo {journal} {Phys. Rev. Lett.}\ }%
  \textbf{\bibinfo {volume} {97}},\ \bibinfo {pages} {180412} (\bibinfo {year}
  {2006})%
  \bibAnnoteFile{NoStop}{barnett06}%
\bibitem{turner07}%
  \BibitemOpen
  \bibfield{author}{%
  \bibinfo {author} {\bibfnamefont{A.~M.}\ \bibnamefont{Turner}}, \bibinfo
  {author} {\bibfnamefont{R.}~\bibnamefont{Barnett}}, \bibinfo {author}
  {\bibfnamefont{E.}~\bibnamefont{Demler}},\ and\ \bibinfo {author}
  {\bibfnamefont{A.}~\bibnamefont{Vishwanath}},\ }%
  \bibfield{journal}{%
  \bibinfo {journal} {Phys. Rev. Lett.}\ }%
  \textbf{\bibinfo {volume} {98}},\ \bibinfo {pages} {190404} (\bibinfo {year}
  {2007})%
  \bibAnnoteFile{NoStop}{turner07}%
\bibitem{song07}%
  \BibitemOpen
  \bibfield{author}{%
  \bibinfo {author} {\bibfnamefont{J.~L.}\ \bibnamefont{Song}}, \bibinfo
  {author} {\bibfnamefont{G.~W.}\ \bibnamefont{Semenoff}},\ and\ \bibinfo
  {author} {\bibfnamefont{F.}~\bibnamefont{Zhou}},\ }%
  \bibfield{journal}{%
  \bibinfo {journal} {Phys. Rev. Lett.}\ }%
  \textbf{\bibinfo {volume} {98}},\ \bibinfo {pages} {160408} (\bibinfo {year}
  {2007})%
  \bibAnnoteFile{NoStop}{song07}%
\bibitem{zhou01}%
  \BibitemOpen
  \bibfield{author}{%
  \bibinfo {author} {\bibfnamefont{F.}~\bibnamefont{Zhou}},\ }%
  \bibfield{journal}{%
  \bibinfo {journal} {Phys. Rev. Lett.}\ }%
  \textbf{\bibinfo {volume} {87}},\ \bibinfo {pages} {080401} (\bibinfo {year}
  {2001})%
  \bibAnnoteFile{NoStop}{zhou01}%
\bibitem{demler02}%
  \BibitemOpen
  \bibfield{author}{%
  \bibinfo {author} {\bibfnamefont{E.}~\bibnamefont{Demler}}\ and\ \bibinfo
  {author} {\bibfnamefont{F.}~\bibnamefont{Zhou}},\ }%
  \bibfield{journal}{%
  \bibinfo {journal} {Phys. Rev. Lett.}\ }%
  \textbf{\bibinfo {volume} {88}},\ \bibinfo {pages} {163001} (\bibinfo {year}
  {2002})%
  \bibAnnoteFile{NoStop}{demler02}%
\bibitem{imambekov03}%
  \BibitemOpen
  \bibfield{author}{%
  \bibinfo {author} {\bibfnamefont{A.}~\bibnamefont{Imambekov}}, \bibinfo
  {author} {\bibfnamefont{M.}~\bibnamefont{Lukin}},\ and\ \bibinfo {author}
  {\bibfnamefont{E.}~\bibnamefont{Demler}},\ }%
  \bibfield{journal}{%
  \bibinfo {journal} {Phys. Rev. A}\ }%
  \textbf{\bibinfo {volume} {68}},\ \bibinfo {pages} {063602} (\bibinfo {year}
  {2003})%
  \bibAnnoteFile{NoStop}{imambekov03}%
\bibitem{bogoliubov47}%
  \BibitemOpen
  \bibfield{author}{%
  \bibinfo {author} {\bibfnamefont{N.~N.}\ \bibnamefont{Bogoliubov}},\ }%
  \bibfield{journal}{%
  \bibinfo {journal} {J. Phys. (USSR)}\ }%
  \textbf{\bibinfo {volume} {11}},\ \bibinfo {pages} {23} (\bibinfo {year}
  {1947})%
  \bibAnnoteFile{NoStop}{bogoliubov47}%
\end{thebibliography}
%

\end{document}